\documentclass[aps,pra,twocolumn,notitlepage,superscriptaddress,floatfix]{revtex4-1}

\usepackage{graphicx}
\usepackage{dcolumn}
\usepackage{bm}
\usepackage{amssymb}
\usepackage{microtype}
\usepackage{xfrac}
\usepackage{makecell}
\usepackage{array}
\usepackage[version=4]{mhchem}
\usepackage{gensymb}
\usepackage{multirow}
\usepackage{physics}
\usepackage[colorlinks=true]{hyperref}
\usepackage{physics}
\usepackage{bm}

\newcommand{\h}{\mathbf h}

\renewcommand\({\begin{equation}}	
\renewcommand\){\end{equation}}

\renewcommand\[{\begin{eqnarray}}	
\renewcommand\]{\end{eqnarray}}

\begin{document}

\title{Zeeman term for the N{\'e}el vector in a two sublattice antiferromagnet using Dzyaloshinsky-Moriya interaction and magnetic field}
\author{Sayak Dasgupta}
\affiliation{Department of Physics and Astronomy $\&$ Stewart Blusson Quantum Matter Institute, University of British Columbia, Vancouver,  British Columbia V6T 1Z1, Canada}
\affiliation{Institute for Solid State Physics, University of Tokyo, Kashiwa 277-8581, Japan}
\author{Ji Zou}
\affiliation{Department of Physics and Astronomy, University of California, Los Angeles, California 90095, USA}

\begin{abstract}
We theoretically investigate the dynamics of solitons in two sublattice antiferromagnets under external perturbations, focusing on the effect of Dzyaloshinsky-Moriya (DM) interactions.  To this end, we construct a micromagnetic field theory for the antiferromagnet in the presence of the external magnetic field, DM interaction, and spin-transfer torque. In particular, we show external magnetic field and spin current couple to N{\'e}el vector in a Zeeman-like manner when DM interactions present, which can be used to efficiently drive antiferromagnetic solitons of different dimensions. Besides, we  study the effect of  straining the local lattice. It can serve as an external handle on the N{\'e}el field inertia and thus dynamical properties. Our findings may find applications in antiferromagnetic spintronics.
\end{abstract}

\maketitle

\section{Introduction}
\label{sec_intro}
Antiferromagnets hold a promise for a faster spintronics platform. The spin wave dynamics of an antiferromagnetic system is controlled by an energy scale $\propto J$, where $J$ is the antiferromagnetic exchange. For ferromagnets the same scale is $\propto \sqrt{K J}$ where $K$ is a local anisotropy. In most materials $J\gg K$. The energy scale for the antiferromagnet translates to a frequency scale of a few THz. Antiferromagnets offer another significant advantage over ferromagnetic devices. Since the net magnetic moment largely cancels over a unit cell, they do not produce stray fields. This is particularly important in device design, where we would like our individual memory components to be isolated from one another. It is in stark contrast to ferromagnets where solitons do not possess an inertia and dynamics is controlled by the local spin Berry phase \cite{nunez2006,RMPYaroslav2018,shick2010,Jungwirth2016,Marrows2016}. 

However, these advantages also present a significant handicap--of coupling antiferromagnetic solitons to external probes. The absence of a local spin density implies a  minor response to spin currents. The response to external magnetic fields is also tuned down by a factor of the exchange strength. One way to manipulate these solitons is to transfer linear momentum, exploiting the inertial dynamics of the solitons \cite{skkim14,tveten14,qaiumzadeh18}. This can be achieved, for instance, by using magnons to scatter from the domain walls. Other methods involve creating a local Berry phase which can then be coupled to an external spin current field. This technique was used in Ref. \cite{dasgupta17} to generate a Magnus force for an antiferromagnetic vortex.

We know from the classic work of Schryer and Walker \cite{schryer74} that,  in a collective coordinate picture \cite{tretiakov08}, an external magnetic field acts like a force on the ferromagnetic domain wall in one dimension. This construction can be extended generically to any spatial dimension. In the ferromagnetic case, the gyroscopic dynamics causes the force to act in the angular momentum channel, leading to a precession of the domain wall.

In antiferromagnets, a local density of magnetization is energetically costly. The dynamics is expressed in terms of soft modes, which are spin configurations with vanishing net spin density. In the case of a two sublattice antiferromagnet, this is the N{\'e}el field. The magnetization density follows the soft mode dynamics and renders an inertial mass to the soft modes. Thus the dynamics in the antiferromagnet is inertial---A force produces a linear acceleration, not a precession\cite{tveten14,dasgupta2020}.

In order to propel antiferromagnetic domain walls easily, one may then hope to use the analogue of the Zeeman field for the N{\'e}el vector. One question naturally arises--what would be equivalent to the magnetic field for the antiferromagnet? This question was addressed by Gomonay $et~al$ \cite{gomonay16} for the two sublattice case. They pointed out that a N{\'e}el spin-orbit field, induced by an electrical current \cite{zelezny14,Wadley2016}, has a Zeeman-like coupling to the N{\'e}el vector (staggered magnetization), which they utilized to drive the one-dimensional domain wall efficiently.

In this paper, we find another situation where  such a Zeeman-like coupling emerges in an antiferromagnet. In particular, we show that the Dzyaloshinsky-Moriya (DM) \cite{dzyaloshinsky58,moriya60} interaction is the key ingredient. The DM interaction creates a local magnetization density which can then respond to both external magnetic fields and spin currents through  Zeeman-like terms. In addition to this, we investigate the effects of straining the local lattice on the staggered magnetization field. The presence of a nonzero strain tensor would modify the inertia of the N{\'e}el field. Thus strain can potentially function as a handle on the dynamics of antiferromagnetic solitons.

Our approach will be that of collective coordinates, developed for describing the slow dynamics of magnetic textures in ferromagnets \cite{tretiakov08} and antiferromagnets \cite{tveten14}. The dynamics of the texture is described through a set of coordinates $q_i$, which represent soft modes of the texture. These are usually restricted to the position and orientations of the soliton. The kinetic energy of an antiferromagnet is expressed as $M_{ij}\dot{q}_i\dot{q}_j$, where $M_{ij}$ is a symmetric inertia tensor. The generalized force conjugate to the coordinate $q_i$ is $F_i = - \partial U /\partial q_i$ with $U$ being the total potential energy. The dissipative force is given by $F_i^v = -D_{ij}\dot{q}_j$. The inertia and dissipation tensors are proportional to each other $D_{ij} = M_{ij}/T$; the relaxation time $T$ is inversely proportional to Gilbert damping constant $\alpha$ \cite{skkim14}. 

Although we use collective coordinates as our degrees of freedom, we shall not use the Landau-Lifshitz equations for the individual sublattices. Instead, we take the micromagnetic field theory picture presented in  Ref. \cite{tveten14,skkim14,dasgupta17} and figure out the potential energies (or gauge theories) that are spawned by adding external perturbations. To facilitate this, we briefly review the micromagnetic field theory for two sublattice antiferromagnets in Sec.~\ref{sec.2_sub_micro}. We then move onto the effects of the individual perturbations: namely a magnetic field, a DM interaction, and a spin-transfer torque in Sec. \ref{sec.external_perturbations}. The meat of our discussion lies in Sec. \ref{sec.combined_int} where we deal with the effect of simultaneous perturbations. This construction is essential for a propulsion mechanism. Finally we gather our results in Sec. \ref{sec.conc}.

\section{Two sublattice micromagnetics}
\label{sec.2_sub_micro}

In this section, we derive the micromagnetic Lagrangian for the two sublattice antiferromagnet along the lines of Ref. \cite{tveten14}. Our description is in terms of the magnetization field represented by the unit vectors $\mathbf{m}(\mathbf{r},t)$. The length of the magnetization, $\mathcal{M}$, is a constant and is connected to the underlying spin density $\mathcal{J}$ through the relation $\mathcal{M} = \gamma \mathcal{J}$ with   gyromagnetic ratio $\gamma$.

For antiferromagnets, each magnetic unit cell comprises two or more magnetization fields $\mathbf{m}_i$ which are constrained by the exchange interaction to follow $\sum_{i}\mathbf{m}_i = 0$. To make this explicit,  we  convert the nearest neighbour exchange into:
\begin{eqnarray}
H_{\text{exchange}} &=& J \sum_{<i,j>} \mathbf{S}_i\cdot\mathbf{S}_j \\ \nonumber
&=& \frac{JS^2}{2}\sum_{\alpha}\left(\sum_{i}\mathbf{m}_i\right)_{\alpha}^2 - \frac{N}{2}\sum_{\alpha} S^2.
\end{eqnarray}
Here $\sum_{i}\mathbf{m}_i$ is a sum over all the spins that constitute the antiferromagnetic unit cell---if there are $N$ sublattices, the sum is over $N$ spins. The other sum $\alpha$ is over the lattice, broken down into the magnetic unit cell clusters. The second term is dropped as it is constant and does not enter equations of motion. 

In general, to get to the continuum model, we express the vector fields $\mathbf{m}_i$ in terms of the appropriate normal modes of the systems, dictated by the point group symmetry of the order, and expand the exchange interaction (and the other energies) in them \cite{dasgupta2020}.

The particular construction of the field theory depends on the specific lattice geometry. However, generically they all stem from labelling the sublattice magnetizations as individual fields and then putting them together by expressing the respective magnetization fields in terms of the normal modes. These are of two kinds---soft modes which do not break the constraint $\sum_{i}\mathbf{m}_i = 0$, and hard modes which do, inducing a net magnetization per unit cell. 

Solitonic dynamics in ferromagnets is dominated by gyroscopic effects generated by the local angular momentum density. Thus, to propel a ferromagnetic vortex in the $x$ direction of the $xy$ plane, one applies a force in the $y$ direction \cite{thiele73}. Similarly, exerting a force to a domain wall in a uniaxial ferromagnet primarily generates a precession about the long axis \cite{schryer74}. To propel it forward, one has to apply a torque to it, for example through the adiabatic spin-transfer torque \cite{bazaliy98,slonczewski02}. This is not the situation in antiferromagnets where a net angular momentum density is usually a secondary effect from local anisotropy and fights with a much larger exchange interaction.

A continuum theory of a collinear antiferromagnet with two sublattices operates with two slowly varying (in space) fields $ \mathcal{M} \mathbf m_1(\mathbf r)$ and $ \mathcal{M} \mathbf m_2(\mathbf r)$. $ \mathcal{M} $ is the moment size and $\mathbf m_1$, $\mathbf m_2$ are unit vector fields. In a state of equilibrium, $\mathbf m_1(\mathbf r) = - \mathbf m_2(\mathbf r)$. More generally, the two sublattice fields are expressed in terms of dominant staggered magnetization $\mathbf n = (\mathbf m_1 - \mathbf m_2)/2$ and small uniform magnetization $\mathbf m = \mathbf m_1 + \mathbf m_2$. The constraints $|\mathbf m_1|^2 = 1$ and $|\mathbf m_2|^2 = 1$ translate into  
\begin{equation}
\mathbf m \cdot \mathbf n = 0, 
\quad
|\mathbf n|^2 = 1 - |\mathbf m|^2/4 \approx 1;
\label{eq:constraints}
\end{equation}
the last approximation is valid as long as $|\mathbf m|^2 \ll 1$.

\begin{figure}[b]
    \includegraphics[width=\columnwidth]{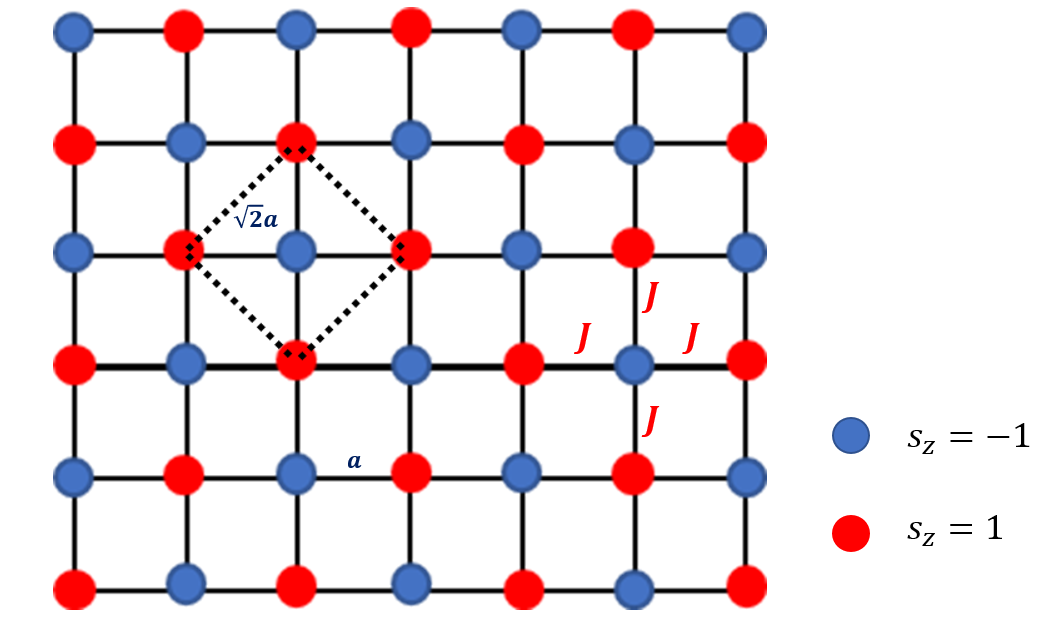}
\caption[Square-lattice antiferromagnet]{This figure shows the two dimensional two sublattice antiferromagnet. Red sites have their spins out of plane and blue spins have spins into the plane. The unit cell for each sublattice is marked in dashed lines. The exchanges are isotropic and are marked.}
\label{fig:square-afm}
\end{figure}

\subsection{The kinetic term and spin wave spectrum}
We demonstrate the calculation of the spin wave spectrum for a two sublattice antiferromagnet on a square lattice of side length $a$. The only interaction present is the nearest neighbour Heisenberg exchange with strength $J$. The kinetic term for the antiferromagnet emerges from the Berry phases of the two sublattice magnetizations $\mathbf{m}_{1,2}$ \cite{haldane86}. The total Berry phase for the unit cell:
\begin{equation}
\label{eq.kin-L-1}
\mathcal{L}_B = \mathcal{J} (\mathbf{a}_{1}.\dot{\mathbf{m}}_{1} + \mathbf{a}_{2}.\dot{\mathbf{m}}_{2}).
\end{equation}
Here $\mathcal{J} = S/(2a^2)$ is the density of angular momentum in two dimensions with $S$ as the moment (spin) length. While choosing the vector potentials $\mathbf{a}_{1,2}$ for the two sublattices,  we adopt different gauges, such that the Dirac string of the two monopoles lie on opposite hemispheres of the magnetization sphere. This ensures that neither $\mathbf{m_{1/2}}$ is near a Dirac string. The convenient choice is $\mathbf{a}_{1}(\mathbf{m})$ $ = \mathbf{a}(\mathbf{m})$ and $\mathbf{a}_{2}(\bf{m})$ $ = \mathbf{a}(-\bf{m})$ \cite{ivanov95,skkim14,dasgupta18}.

In the equilibrium state when $\bf{m}_{1} = - \bf{m}_{2}$, the Berry phases of the two sublattices cancel exactly. This can be seen for the standard gauge choice of the vector potential $a_{\theta} = 0$ and $a_{\phi} =(\cos\theta \pm 1)/\sin\theta$. The Dirac string carries a `flux' of $+4\pi$ either through the north or south pole. If we put the string through the south pole for $\mathbf{m}_1$ and through the north pole for $\mathbf{m}_2$ we have in equilibrium $\mathcal{L}_B = \mathcal J \left[\mathbf{a}(\mathbf{n}) - \mathbf{a}(-(-\mathbf{n}))\right]\cdot\dot{\mathbf{n}} = 0$.

The lowest non-vanishing kinetic terms are obtained by expanding the vector potentials using $|\mathbf{m}|$ as a small parameter. Individually, $\mathbf{a}_{1}\cdot\dot{\mathbf{m}}_{1} = \mathbf{a}_{1}(\mathbf{m/2 + n})\cdot(\dot{\mathbf{m}}/2 + \dot{\mathbf{n}})$ and $\mathbf{a}_{2}\cdot\dot{\mathbf{m}}_{2} = \mathbf{a}_{2}(\mathbf{m/2 - n})\cdot(\dot{\mathbf{m}}/2 - \dot{\mathbf{n}})$ . Expanding to quadratic order in $|\mathbf{m}|$ and $|\dot{\mathbf{n}}|$, the kinetic term Eq.~(\ref{eq.kin-L-1}) yields the following:
\begin{eqnarray}
\label{eq.Berry_terms_Lag}
\mathcal{L}_B/\mathcal{J} &=&\left[\mathbf{a}_1(\mathbf{n}) + \mathbf{a}_2(-\mathbf{n})\right]\cdot\frac{\dot{\mathbf{m}}}{2} \\ \nonumber
&+& \left[\mathbf{a}_1(\mathbf{n}) - \mathbf{a}_2(-\mathbf{n})\right]\cdot\dot{\mathbf{n}} \\ \nonumber
&+& \frac{m_{i}}{2}\left[\frac{\partial\mathbf{a}_1(\mathbf{n})}{\partial n_{i}} - \frac{\partial\mathbf{a}_2(-\mathbf{n})}{\partial n_{i}}\right]\cdot\frac{\dot{\mathbf{m}}}{2} \\ \nonumber
&+&\frac{m_{i}}{2}\left[\frac{\partial\mathbf{a}_1(\mathbf{n})}{\partial n_{i}} + \frac{\partial\mathbf{a}_2(-\mathbf{n})}{\partial n_{i}}\right]\cdot\dot{\mathbf{n}}
\end{eqnarray}
We have the identity $\partial_{n_i}\mathbf{a}_1(\mathbf{n}) - \partial_{n_i}\mathbf{a}_2(-\mathbf{n})=0$, from the definition of the vector potentials. This cancels the second and third terms.  In the first term, we now transfer the time derivative to $\mathbf{a}$ using an integration by parts and combine with the corresponding vector potential term from the last line to get: 
\begin{equation}
m_i\dot{n}_k\left[\frac{\partial a_k(\mathbf{n})}{\partial n_{i}} - \frac{\partial a_i(\mathbf{n})}{\partial n_{k}}\right] = \dot{\mathbf{n}}\cdot(\mathbf{n}\times\mathbf{m}),
\end{equation}
where we have used $\bm{\nabla}_{\mathbf{n}}\times\mathbf{a} = -\mathbf{n}$.

The potential energy is obtained from the Heisenberg exchange:
\begin{eqnarray}
   U &=& JS^2\sum_{\langle i,j\rangle} \mathbf{m}_i\cdot\mathbf{m}_j, \\ \nonumber 
   &=& \frac{JS^{2}}{2}\sum_{\alpha}(\mathbf{m}_1 + \mathbf{m}_2)_{\alpha}^2 \\ \nonumber
     &=& \int dV~\frac{JS^{2}}{2} \left[\frac{2\mathbf{m}^2}{a^2} + (\partial_i\mathbf{n})^2 + \frac{(\partial_i\mathbf{m})^2}{2}\right],
\end{eqnarray}
where $J$ is the Heisenberg exchange strength and in the second line we have dropped the constant term. In the second line, we have expressed the summation over nearest neighbours in terms of summation over two site magnetic unit cells $\alpha$. We can see that the uniform magnetization picks up an energy contribution from the exchange interaction at the zeroth order in gradients and is hence a hard mode. The N{\'e}el field $\mathbf{n}$ only appears through gradients and is the typical example of a soft mode in antiferromagnetic systems.

\subsection{Spin Waves}
The procedure to obtain the effective spin wave field theory is similar to the planar ferromagnet \cite{dasgupta20}: we integrate out the hard field and express the theory in terms of the soft field. This process generates an inertia for the soft mode. Since $\mathbf{m}$ is hard, we shall drop its gradient terms. Let us carry this out explicitly:
\begin{equation}
\label{eq.n.sw}
    \mathcal{L} = \frac{S}{2a^2}~\dot{\mathbf n}\cdot(\mathbf n \times \mathbf m) - \frac{JS^{2}}{2} \left[\frac{2\mathbf{m}^2}{a^2} + (\partial_i\mathbf{n})^2 \right].
\end{equation}
Now we can solve for the hard field $\mathbf{m} = (\dot{\mathbf n}\times\mathbf n)/(4 J S)$, implying $\vb{m}$ is a slave variable to the N{\'e}el field $\vb{n}$ in this treatment. Substituting this solution back into the Lagrangian, we obtain a field theory for the soft N{\'e}el field:
\begin{equation}
\label{eq.L-n-original}
    \mathcal{L} = \frac{\rho}{2}\dot{\mathbf n}^{2} -\frac{JS^{2}}{2}\left(\bm{\nabla}\mathbf n\right)^2,
\end{equation}
with $\rho = 1/(8Ja^2)$. Here we have used $(\dot{\mathbf{n}}\times\mathbf{n})^2 = \dot{\mathbf{n}}^2$ as $\mathbf{n}\cdot\dot{\mathbf{n}} = 0$, following from the unit vector constraint of $\mathbf{n}$.

The ordered ground state $\vb{n}_0(\theta,\phi)$ spontaneously breaks the $SO(3)$ symmetry of the system up to $SO(2)$.
 Hence in this case, there are two Goldstone modes, residing in the coset space $S^2=SO(3)/SO(2)$, one for each continuous degree of freedom, dispersing linearly according to $\omega = c k$, with $c = \pm (2\sqrt{2}JSa) $. They classically correspond to the opposite circular polarizations of the small-angle oscillations of $\delta\vb{n}\perp \vb{n}_0$.

\subsection{Strain}
The strain to moment coupling is expressed through the energy density \cite{tchernyshyov02}:
\begin{equation}
\mathcal{U}_{ME} = S^2\sum_{ij}\left[\frac{\partial J(\mathbf{u})}{\partial \mathbf{u}_{ab}}\cdot\delta\mathbf{u}_{ab}\right]\mathbf{m}_i\cdot\mathbf{m}_j, 
\end{equation}
where $\mathbf{u}_{ab} = \mathbf{u}_a - \mathbf{u}_b$ with $\mathbf{u}$ as the lattice displacement field. On the nearest neighbour square lattice, the only strain components that couple to the Heisenberg Hamiltonian are $\epsilon_{xx}$ and $\epsilon_{yy}$, where $\epsilon_{ij} =  (\partial_i u_j + \partial_j u_i)/2$ is the linear strain tensor. If the system has next-nearest neighbour interactions, we can couple to those using the off-diagonal strain $\epsilon_{xy}$.  The off diagonal strain will appear in two dimensions for non collinear magnetic ordering, for instance the Mn$_3$X group of 120$^\circ$ ordered antiferromagnets \cite{cable1993,chen2020,soh2020}.

\begin{figure}[t]
    \includegraphics[width=\columnwidth]{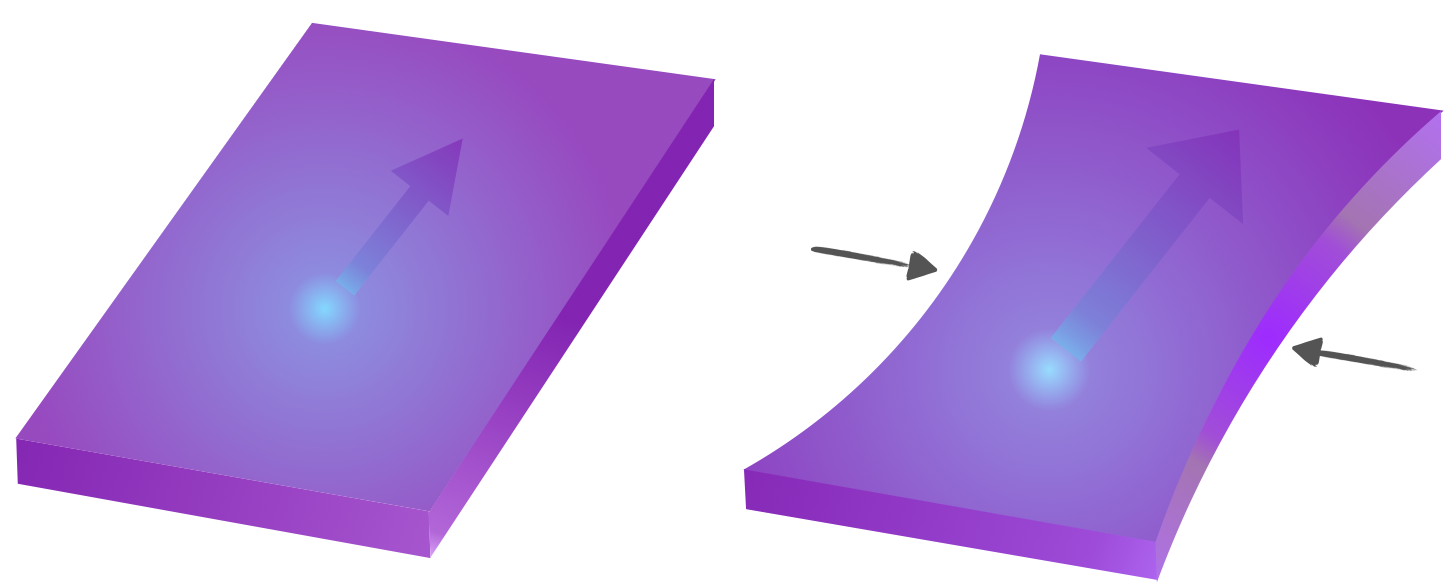}
\caption{Schematic of strain serving as a handle on dynamics of solitons in two dimensions. The strain  modifies the inertia associated with N{\'e}el field. It can also introduce anisotropies, leading to anisotropic spin wave velocity.}
\label{strain}
\end{figure}

To lowest order in gradients, the strain couples to the uniform magnetization $\mathbf{m}$ and gradients of the N{\'e}el vector $\partial_i \mathbf{n}$. The dominant effect is through a coupling to the uniform magnetization $\mathbf{m}$. This produces an energy density:
\begin{equation}
    \mathcal{U}_{ME} = J'S^{2}\frac{(\epsilon_{xx} + \epsilon_{yy})\mathbf{m}^2}{2 a^2},
\end{equation}
where $J' = (\partial_i J)$ and we have assumed $\partial_x J = \partial_y J$ from the local cubic symmetry. This modifies the inertia for the N{\'e}el field:
\begin{equation}
    \frac{1}{\rho'} = \frac{1}{\rho} \left[ 1 + \frac{J'(\epsilon_{xx} + \epsilon_{yy})}{2J} \right].
\end{equation}
It  serves as an external handle on the N{\'e}el field inertia which can be exploited to control its dynamical properties, especially in the case of solitons (see Fig.~\ref{strain}). This presents a new avenue to manipulate the frequency response for two sublattice antiferromagnets.

The next higher order coupling is to the gradients of the soft N{\'e}el field. This coupling modifies the spin wave velocity and makes it anisotropic. This is expected since strains induce an additional two fold anisotropy in the plane. The velocities are now given by: 
\begin{equation}
    \mathbf{c} = c \left( 1 + \frac{3J'}{2J} \epsilon_{xx} + \frac{J'}{2J} \epsilon_{yy}, 1 + \frac{3J'}{2J} \epsilon_{yy} + \frac{J'}{2J} \epsilon_{xx}\right),
\end{equation}
with $c = \pm (2\sqrt{2}JSa)$.

\subsection{Solitons}
We are interested in the situations where the only spatial dependence of the staggered magnetization field $\mathbf{n}$ is at the location of topological defects. These regions are characterized by a skyrmion density defined using the N{\'e}el vector field:
\begin{equation}
    N_{sk} = \frac{1}{4\pi}\int dq_i dq_j~\mathbf{n} \cdot \left(\frac{\partial \mathbf{n}}{\partial q_i} \times \frac{\partial \mathbf{n}}{\partial q_j}\right).
\end{equation}
Here $(q_i,q_j)$ are collective coordinates conjugate to each other. Typical examples for the two sublattice case are---domain walls characterized by the conjugate set of location and orientation of the domain wall $(Z,\Phi)$, and the vortex with it core center $(X,Y)$ serving as the conjugate set. 

\begin{figure}[t]
\includegraphics[width=\columnwidth]{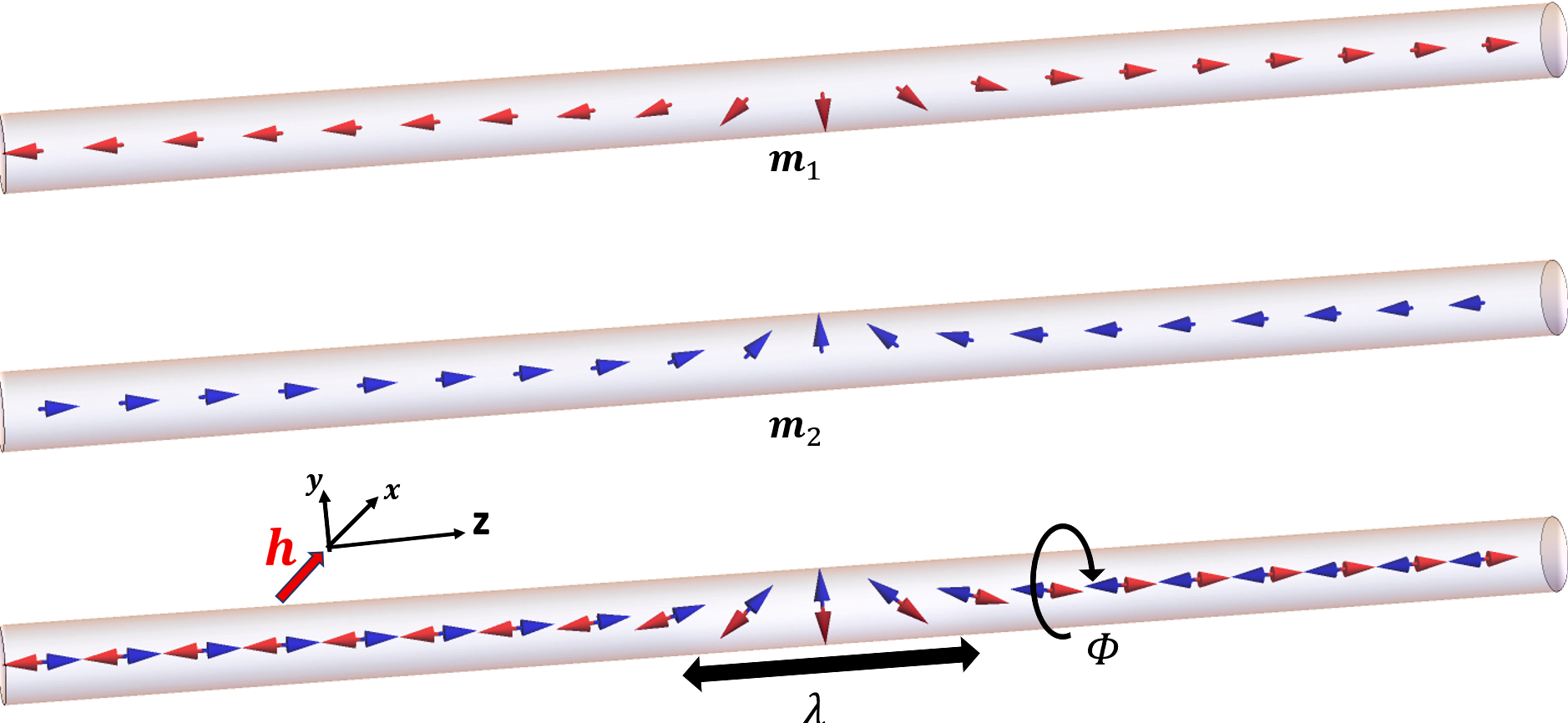}
\caption[Uniaxial two sublattice antiferromagnet]{On the top we show the constituent sublattice magnetizations $\mathbf{m}_{1,2}$. These sublattices combine to form the antiferromagnet. A typical soliton in one dimension is a domain wall shown on the bottom. The domain wall is a soliton interpolating between the two unidirectional ground states of the one dimensional antiferromagnet.}
\label{fig:domai-wall-afm}
\end{figure}

\textit{Uniaxial domain wall}: The uniaxial domain wall is produced by an easy axis anisotropy. Choosing this to lie along the $z$ axis we get:
\begin{equation}
\label{eq.energy-density}
\mathcal U(\mathbf n) =  
		\frac{A}{2} \left|\frac{\partial \mathbf n}{\partial z}\right|^2
		+ \frac{K}{2} |\mathbf e_3 \times \mathbf n|^2.
\end{equation}
Here $A>0$ characterizes the strength of exchange, $K>0$ is the easy axis anisotropy, and $\mathbf e_3 = (0,0,1)$. This system has two uniform ground states $\mathbf n = \pm \mathbf e_3$, linear excitations in the form of spin waves with the dispersion $\omega^2 = (K + Ak^2)/\rho$, and nonlinear solitons in the form of domain walls which interpolate between the two ground states. Static domain walls in $\mathbf{n} = (\sin\theta(z)\cos\phi,\sin\theta(z)\sin\phi,\cos\theta(z))$ have width $\lambda = \sqrt{A/K}$ and are parametrized in spherical angles $\theta(z)$ and $\phi(z)$ as follows: 
\begin{equation}
\cos{\theta(z)} = \pm \tanh{\frac{z-Z}{\lambda}}, 
\quad 
\phi(z) = \Phi. 
\label{eq:domain-wall}
\end{equation}
Position $Z$ and azimuthal angle $\Phi$ represent the two zero modes of the system associated with the global symmetries of translation and rotation see Fig.~\ref{fig:domai-wall-afm}. Weak or local external perturbations do not alter the shape of the soliton significantly and mostly induce the dynamics of $Z$ and $\Phi$.

\textit{Planar vortex} : This topological feature is stable in two spatial dimensions with an easy plane anisotropy, $K<0$ in Eq.~(\ref{eq.energy-density}). The uniform ground states are $\mathbf{n} = (\cos\phi, \sin\phi, 0)$. A vortex centred at the origin is parametrized as:
\begin{equation}
    \label{eq.vortex}
    e^{i \phi(\mathbf{r})} = \left( \frac{x + iy}{|x + iy|}\right)^n, ~~ \cos\theta(\mathbf{r}) = \pm f_n (r/\lambda).
\end{equation}
Here $n \in \mathbb{Z}$ is the vortex winding number. The magnetization leaves the plane at the cores and this is captured by the function $f(\zeta)$ with $f_n(0) = 1$ and $f_{n}(\infty) = 0$. The core size is controlled by the same length scale as the domain wall, $\lambda = \sqrt{A/|K|}$. 
\section{External Perturbations}
\label{sec.external_perturbations}
We now consider the situation where the only spatial dependence of the staggered magnetization field $\mathbf{n}$ is at the location of defects. The theory we work with is
\begin{equation}
\mathcal{L} = \mathcal{J}\dot{\mathbf n}\cdot(\mathbf n \times \mathbf m)-  \left(\frac{\mathcal{M}^2}{2\chi}\right)\mathbf{m}^{2} - \mathcal{U}_{\text{ext}}[\bm{\zeta},\mathbf{n},\mathbf{m}],
\end{equation}
where we have absorbed the Heisenberg exchange strength into a spin susceptibility $\chi$. $\bm{\zeta}$ in the theory is an external (pseudo)vector field (it can be a general tensor field, such as the strain tensor we have discussed). Our main objective is to see how $\bm{\zeta}$ modifies the Lagrangian density, in particular how it couples to the soft mode $\mathbf{n}$. Once we have an understanding of these couplings, we can study their effects on solitons in the staggered magnetization order, such as uniaxial domain walls and planar vortices. We outline the manner in which these solitons can be effectively moved in space by coupling to the order parameter. 

These external vector fields couple either to the uniform magnetization $\mathbf{m}(\mathbf{r},t)$ or the staggered magnetization $\mathbf{n}(\mathbf{r},t)$ in the Lagrangian. This is broadly guided by symmetries like time reversal and mirror planes of the spin Hamiltonian. Fields, which couple to $\mathbf{m}$, produce a gauge coupling to $\dot{\mathbf{n}}$, on integrating out $\mathbf{m}$. This is the case with perturbations like an external magnetic field $\mathbf{h}(\mathbf{r},t)$ or a spin transfer torque characterized by the electron drift velocity $\mathbf{u}(\mathbf{r},t)$. Such terms require a spatial or temporal variation of the external vector field to produce solitonic motion \cite{dasgupta17,yamane17}.

The coupling to $\mathbf{n}$ gives rise to terms like $(A_{ij}\zeta_in_j)^n$, where $n = 1,2$ is the cases we study. Here $\zeta_i$ represents an external field sourced from a combination of terms like the Dzyaloshinski-Moriya interaction, external magnetic fields, or combinations. This term acts as a potential energy density which can generate a force (or torque) on a soliton. Note here, that an antiferromagnetic soliton by virtue of Eq.~(\ref{eq.L-n-original}) is inertial, i.e. a force propels an antiferromagnetic domain wall instead of making it precess. We show that Dzyaloshinski-Moriya interactions generate such terms and can be used to propel solitons.

In the course of working out these contributions to the energy density, one particularly useful identity we repeatedly use is:
\begin{equation}
    (\mathbf{\dot{n}} \times \mathbf{n})\cdot(\mathbf{A}\times\mathbf{n}) = \mathbf{A}\cdot[\mathbf{n}\times (\mathbf{\dot{n}} \times \mathbf{n})] = \mathbf{A}\cdot\mathbf{\dot{n}}.
\end{equation}
This identity follows in a straightforward manner from $\mathbf{n}^2 = 1 - (\mathbf{m}^2/4) \simeq 1$. Corrections to this assumption modify the inertia $\rho$. Since $\mathbf{A}$ in our theory is already a perturbation, these are higher order corrections and will be suppressed.

\subsection{Magnetic field}
The external magnetic field couples to the uniform magnetization $\mathcal{U} = -\mathcal{M}\mathbf{h}\cdot\mathbf{m}$ to form a Zeeman term. This adds to  the Lagrangian density:
\begin{equation}
\label{eq:L-n-m}
\mathcal{L}[\mathbf{m},\mathbf{n}] = \mathcal{J}\dot{\mathbf{n}}\cdot(\mathbf{n}\times \mathbf{m}) -  \left(\frac{\mathcal{M}^2}{2\chi}\right)|\mathbf{m}|^{2} +  \mathcal{M}\mathbf{h}\cdot\mathbf{m}.
\end{equation}
A straightforward minimization with respect to $\mathbf{m}$ gives $\mathbf{m} = \chi \mathcal{J}\dot{\mathbf{n}}\times\mathbf{n}/ \mathcal{M}^2 + \chi \mathbf{h}/\mathcal{M}$, which violates the constraint $\mathbf{m}\cdot\mathbf{n} = 0$. To ensure the perpendicularity, we resolve $\mathbf{h}$ into a component perpendicular to $\mathbf{n}$, $\mathbf{h}_{\perp} = \mathbf{n}\times(\mathbf{h}\times\mathbf{n})$ which enters the Zeeman coupling $\mathbf{m}\cdot\mathbf{h}_{\perp}$ to produce a term $(\mathbf{n}\times\mathbf{H})\cdot(\mathbf{n}\times\mathbf{m})$.

Now on solving for $\mathbf{m}$, we obtain $\mathbf{m} =\chi \mathcal{J}(\dot{\mathbf{n}}\times\mathbf{n})/\mathcal{M}^2 + \chi(\mathbf{n}\times\mathbf{h})\times\mathbf{n}/\mathcal{M}$. Substituting this into the Lagrangian we obtain:
\begin{equation}
\mathcal L(\mathbf n) = 
	\frac{\rho (\dot{\mathbf n} - \gamma \mathbf h \times \mathbf n)^2}{2} , 
\label{eq:L-n-h}
\end{equation} 
with the inertia $\rho = \chi/\gamma^2$. The Lagrangian is identical to that of a particle in a rotating frame with an angular velocity $\gamma |\mathbf h|$, causing a texture in $\mathbf{n}$ to precess. There is an additional contribution to the energy in the form of $\mathcal{U}_{H} = - \rho |\gamma \vb{h}\times \vb{n}|^2/2$, which adds to the crystal anisotropy term in the energy functional and resembles the potential energy that leads to the centrifugal force in the rotating frame.

Let us take a closer look at each of the terms in Eq.~(\ref{eq:L-n-h}). The term $\rho \dot{\mathbf n}^2/2$ is the kinetic energy of staggered magnetization, which endows antiferromagnetic solitons with an inertial mass. Supposing a soliton is parametrized by a set of collective coordinates $\mathbf q = \{q_1, q_2, \ldots\}$ such as the position of a domain wall, the coordinates of a vortex core etc., the variation of $\mathbf n$ in time is mediated by the change of these collective coordinates: $\dot{\mathbf n} = \dot{q}_i \partial_{q_i}\vb{n}$. The soliton's kinetic energy is then $M_{ij} \dot{q}_i \dot{q}_j/2$, where $M_{ij} = \rho \int dV \, \partial_{q_i}\vb{n} \cdot \partial_{q_j}\vb{n}$ is the inertia tensor \cite{tveten13}. 

The potential term $\rho |\gamma\mathbf h \times \mathbf n|^2/2$ in Eq.~(\ref{eq:L-n-h}) expresses local anisotropy favouring the direction of $\mathbf n$ orthogonal to the effective field $\mathbf h$. This term modifies the potential landscape $U(\mathbf q)$ of a soliton: 
\begin{equation}
U[\mathbf q, \mathbf h(\mathbf r)] = U[\mathbf q, 0] 
	-  \int dV \, \frac{\rho|\gamma \mathbf h \times \mathbf n|^2}{2}.
\label{eq:U-h}
\end{equation}
To get an idea of what kind of anisotropy this term induces, let us take a look at the energy density for the uniaxial domain wall in Eq.(\ref{eq:domain-wall}) with the easy axis along $\mathbf{\hat{z}}$ as shown in Fig.\ref{fig:domai-wall-afm}. We now introduce a magnetic field $\mathbf{h} = h_0 (\cos\varphi,\sin\varphi,0)$ modifying the energy density:
\begin{eqnarray}
   \mathcal{U} &=& -\frac{K}{2} \cos^2\theta - \frac{\rho\gamma^2}{2}(\mathbf{n}\times\mathbf{h})^2, \\ \nonumber
               &=& -\frac{K}{2}\cos^2\theta - \frac{\rho(\gamma h_0)^2}{2} [\cos^2\theta + \sin^2\theta\sin^2(\phi - \varphi)],
\end{eqnarray}
with $K>0$.

The magnetic field chooses the azimuthal plane for the N{\'e}el domain wall and hence acts as an angle-selector. For a particular direction of the field $(\cos\theta)$ the minimum energy occurs when $|\phi - \varphi |= \pi/2$. In the figure (Fig.\ref{fig:domai-wall-afm}) we point the magnetic field along $\mathbf{\hat{x}}$ which prefers a N{\'e}el wall in the $yz$ plane. The easy axis anisotropy, however, is unaffected in this configuration. This leaves the soliton size unchanged.

To modulate the size of the soliton $\lambda = \sqrt{A/K}$ we need to apply a field along the easy axis $\mathbf{h} = h_0\mathbf{\hat{z}}$. In this configuration the anisotropy $K$ defined in Eq.~(\ref{eq:domain-wall}) is modified to $K \to \Tilde{K}= K - \rho\gamma^2\h_0^2 $. Now for the easy axis scenario since $K > 0$ this leads to an expansion, while for the easy plane scenario where $K<0$ this leads to a constriction of $\lambda$. Thus the magnetic field breaks the $SO(3)$ symmetry of the N{\'e}el vector and allows an external control of the soliton size. 

We remark that, in the easy axis case, the soliton profile is no longer stable when $\Tilde{K}\rightarrow 0$ as the applied magnetic field increases; the system undergoes a spin-flop transition into a spin-flop phase, where the N{\'e}el vector lies within the plane perpendicular to the magnetic field. In the easy plane case, one can utilize the magnetic field to enhance the easy-plane anisotropy, which is essential for the conservation of spin winding and thus is applicable in energy storages \cite{YaroslavPRL2018, jones2020} and related transport experiments~\cite{KimSci2021, Zou2019, YaroslavPRB2020}. 

The cross term $\rho \gamma \mathbf h \cdot(\dot{\mathbf n} \times \mathbf n)$ in Eq.~(\ref{eq:L-n-h}) is linear in time derivative $\dot{\mathbf n}$ and thus quantifies the effective geometric phase for the dynamics of staggered magnetization. This is analogous to the   Coriolis effect in a rotating frame. In the Lagrangian of a soliton, it turns into $A_i \dot{q}_i$, a coupling to an external gauge field 
\begin{equation}
A_i(\mathbf q) = \int dV \, \rho \gamma \mathbf h \cdot 
	\left(
		\frac{\partial \mathbf n}{\partial q_i} \times \mathbf n
	\right).
\label{eq:A} 
\end{equation}

The equations of motion for an antiferromagnetic soliton have the form of Newton's second law for a particle of unit electric charge in this gauge field: 
\begin{equation}
M_{ij} \ddot{q}_j = - \partial U/\partial q_i + E_i + F_{ij} \dot{q}_i - M_{ij} \dot{q}_j/T.
\label{eq:2nd-law}
\end{equation}
The ``magnetic field'' $F_{ij} = -F_{ji}$ is the curl of the gauge potential: 
\begin{equation}
F_{ij} = \frac{\partial A_j}{\partial q_i} - \frac{\partial A_i}{\partial q_j}
	= - 2 \int dV \, \rho \gamma \mathbf h \cdot 
		\left(
			\frac{\partial \mathbf n}{\partial q_i} 
			\times 
			\frac{\partial \mathbf n}{\partial q_j}
		\right).
\label{eq:F-def}
\end{equation}
The ``electric field" 
\( E_i=-\int dV\, \rho \gamma \dot{\vb{h}} \cdot \left( \pdv{\vb{n}}{q_i} \times \vb{n}\right),  \)
arises when $\vb{h}$ depends on time explicitly. 

\subsection{Dzyaloshinski-Moriya Interaction}
We  now examine the effect of adding the antisymmetric exchange or DM interaction \cite{dzyaloshinsky58, moriya60} to the Lagrangian. This interaction exists in an antiferromagnet with broken inversion symmetry intrinsically or at interfaces like sample edges and extended domain walls. It is characterized by the energy density $\mathcal{U}_{\text{DMI}} = \mathbf{D}\cdot(\mathbf{S}_i\times\mathbf{S}_j) = \mathcal{J}^2\mathbf{D}\cdot(\mathbf{m}_i\times\mathbf{m}_j)$ where the direction of the DM vector $\mathbf{D}$ is given by the Moriya rules~\cite{moriya60}.

Their net effect is to induce a weak ferromagnetism in the material, which then couples to external torques and fields. In the presence of a homogeneous DM interaction, the theory takes the form:
\begin{equation}
\label{eq:L-n-m-D}
\mathcal{L} = \mathcal{J}\dot{\mathbf{n}}\cdot(\mathbf{n}\times \mathbf{m}) - \left(\frac{\mathcal{M}^2}{2\chi}\right)|\mathbf{m}|^{2} - \mathcal{J}^2\mathbf{D}\cdot(\mathbf{n}\times\mathbf{m}).
\end{equation}
This adds an extra term to the solution for the staggered magnetization $\mathbf{m} = \chi \mathcal{J}\dot{\mathbf{n}}\times\mathbf{n}/\mathcal{M}^2 - \chi \mathbf{D}\times\mathbf{n}/\gamma^2$. On integrating out the uniform magnetization we obtain:
\begin{equation}
\label{eq.L-n-D}
\mathcal{L} = \frac{\rho(\dot{\mathbf{n}} - \mathcal{J}\mathbf{D})^2}{2}.
\end{equation}
Note that here, unlike in the case of the external magnetic field, there is no additional anisotropy induced by the DM vector. The Lagrangian suggests a steady-state translation for the N{\'e}el soliton $\dot{\mathbf{n}}' \equiv \dot{\mathbf{n}} - \mathcal{J}\mathbf{D}$ with a velocity $\mathbf{v} = \mathcal{J}\mathbf{D}$. In other words it acts as a potential for $\dot{\mathbf{n}}$.
 
 The cross term with the kinetic term gives rise to a vector potential of the form:
\begin{equation}
 A_{i}^{DM} =- \rho\mathcal{J}\int dV \frac{\partial \mathbf{n}}{\partial q_{i}}\cdot\mathbf{D}.
\end{equation}
For the material bulk where the DM vector is a constant, this does not produce an electromagnetic field density $F_{ij}$. However, there are two situations where an exception occurs. One is when $\partial_i\partial_j\mathbf{n} - \partial_j\partial_i\mathbf{n} \neq 0$ as in the case of the antiferromagnetic vortex core where $\mathbf{n}$ is singular \cite{dasgupta20}. In this case,  the vector potential $A_i$ yields a density $F_{XY} = (-2\pi \tilde{n} \rho \mathcal{J})\mathbf{e}_\phi \cdot \mathbf{D}$. Here $\tilde{n}$ is the vorticity density and $\hat{\mathbf{e}}_\phi$ is the azimuthal unit vector.

It is unlikely that this effect is finite in the two-sublattice case as the DM vector tends to point out of the plane. However, it might be present in non collinear antiferromagnets like Mn$_3$Ge.  The other situation occurs at interfaces where the DM vector can become space dependent. In that case, the electromagnetic tensor strength is given by $F_{ij} = \rho \mathcal{J}(\partial_j \mathbf{n} \cdot \partial_i \mathbf{D} - \partial_i \mathbf{n} \cdot \partial_j \mathbf{D})$.

\subsection{Spin-Transfer Torque}
For metallic antiferromagnets, we can transfer angular momentum to each individual sublattice through a conduction band electron current \cite{cheng2014}. The local magnetic moments couple to the electron spins through an s-d exchange\cite{shulei2009}. The coupling polarizes the conduction band to follow the orientation of spins on individual sublattices. This mechanism gives rise to the adiabatic spin transfer torque.

For the ferromagnet, the adiabatic spin transfer torque modifies the time derivative in the Landau-Lifshitz equation to a convective derivative $\partial_t \to \partial_t + \mathbf{u}\cdot\bm{\nabla}$ \cite{tatara08}. Here $\mathbf{u}$ is the drift velocity of electrons related to the electric current $\mathbf{j} = en\mathbf{u}$---with $n$ as the concentration of electrons. 

This manipulation can be extended to the two sublattice antiferromagnet \cite{swaving11}. The kinetic term: 
\begin{equation}
    \mathbf{m}\cdot(\dot{\mathbf{n}} \times \mathbf{n}) \to \mathbf{m}\cdot[(\partial_t + \mathbf{u}\cdot\bm{\nabla})\mathbf{n} \times \mathbf{n}].
\end{equation}
This correction modifies the induced magnetic moment $\mathbf{m} = (\chi \mathcal{J}/\mathcal{M}^2)(\dot{\mathbf{n}} + \mathbf{u}\cdot\bm{\nabla}\mathbf{n})\times\mathbf{n}$, which suggests that nonuniform N{\'e}el fields will induce a magnetization in the presence of a spin current. The Lagrangian reads:
\begin{equation}
\label{eq.L-n-u}
\mathcal{L} = \frac{\rho (\dot{\mathbf{n}} + \mathbf{u}\cdot\bm{\nabla}\mathbf{n})^2}{2}.
\end{equation}
The most immediate effect of this coupling is to modify the spin wave velocities. Comparing this with Eq.~(\ref{eq.n.sw}), we can see that the potential energy density is now:
\begin{equation}
    \mathcal{U} = \frac{JS^2}{2}(\bm{\nabla}\mathbf{n})^2 - \frac{\rho}{2}(\mathbf{u}\cdot\bm{\nabla}\mathbf{n})^2.
\end{equation}
Consider an adiabatic spin current of the form $\mathbf{u} = (u_x,0)$. This modifies the spin wave velocity in the $\hat{\mathbf{x}}$ direction to $c_x = c [1 - (u_x/(2 c^2))]$, where $c = 2\sqrt{2}JSa$. Thus for a generic current direction the spin wave will no longer be isotropic in the plane and will get corrections of the order of $|\mathbf{u}|/c^2$. This, along with strain can be used to modify spin wave magnitudes and polarizations in the two sublattice antiferromagnet.

The adiabatic spin transfer torque needs a local Berry phase density to effect propulsion of a soliton. This implies that the spin transfer torque needs to be applied in addition to a perturbation that creates a local magnetization density to propel an antiferromagnetic soliton. For instance, in Dasgupta $et~al$ \cite{dasgupta17} an external magnetic field was used to generate a local Berry phase density. This coupled to the spin transfer torque to produce a Magnus force for the antiferromagnetic vortex.

\section{Combined Interactions}
\label{sec.combined_int}
Single perturbations couple to the N{\'e}el field in Eq.~(\ref{eq:L-n-h}), Eq.~(\ref{eq.L-n-D}), and Eq.~(\ref{eq.L-n-u}) through $\dot{\mathbf{n}}$. This gives rise to vector potentials. Under certain circumstances where the perturbation is itself nonuniform in time or space, this leads to a finite electromagnetic tensor. However, as shown in \cite{dasgupta17}, a perturbation that is nonuniform in time does not produce a net propulsion of a soliton. Spatially nonuniform magnetic fields do seem to produce a propulsion \cite{Yuan2018,yamane17}.

A better alternative for antiferromagnetic solitons is to use a combination of two (or more) perturbations. This is the situation which we now turn to. The theme of two of these combinations is similar. If we have a magnetic field $\mathbf{h}(\mathbf{r},t)$ or a DM interaction $\mathbf{D}(\mathbf{r},t)$ locally (at the location of the soliton) inducing a small magnetic moment which the spin current $\mathbf{u}(\mathbf{r},t)$, latches on to and generates a displacement of the soliton. The other combination, a DM interaction and an external magnetic field, will lead to an energy density which we show is structurally identical to N{\'e}el spin orbit torque used in \cite{gomonay16}.

\subsection{DM interaction and external magnetic field}
If these two types of terms are simultaneously present in the system, the Lagrangian density takes the form:
\begin{equation}
\label{eq.L-n-D-H}
\mathcal{L} = \frac{\rho\left[\dot{\mathbf{n}} + \gamma (\mathbf{n}\times\mathbf{h}) - (\mathcal{M}/\gamma)\mathbf{D}\right]^2}{2} - \mathcal{U}(\mathbf{n},\mathbf{D},\mathbf{h}).
\end{equation}
The cross term of interest is:
\begin{equation}
\label{eq.cross-dm-h}
\mathcal{U}_{\text{DM-h}} = -\rho\mathcal{M}~\mathbf{n}\cdot(\mathbf{D}\times\mathbf{h}).
\end{equation}
This term acts as a `Zeeman' term but for the staggered magnetization with an effective magnetic field $\mathbf{h}_{\text{eff}} = (\mathbf{D}\times\mathbf{h})$. Note that in the presence of a DM interaction the extra uniform magnetization that is induced is $\mathbf{m} \propto (\mathbf{D}\times\mathbf{n})$. It is this extra induced ferromagnetic moment that `Zeeman' couples with the external magnetic field. This coupling has been previously studied in the context of weak ferromagnets, for a review see \cite{Bar_yakhtar_1985}.

Here we shall look at it in the context of the two sublattice antiferromagnet, cast in the collective coordinate scheme. This achieves two goals---firstly, it becomes clear that the term is a force on an massive particle (the soliton). Secondly, once we have the collective coordinate scheme set up we can quickly determine the effect of this term on the dynamics of the regular solitons--- uniaxial domain wall, vortex, skyrmion, and hedgehog. This analysis has so far not been presented in the literature.

\begin{figure}[t]
    \includegraphics[width=\columnwidth]{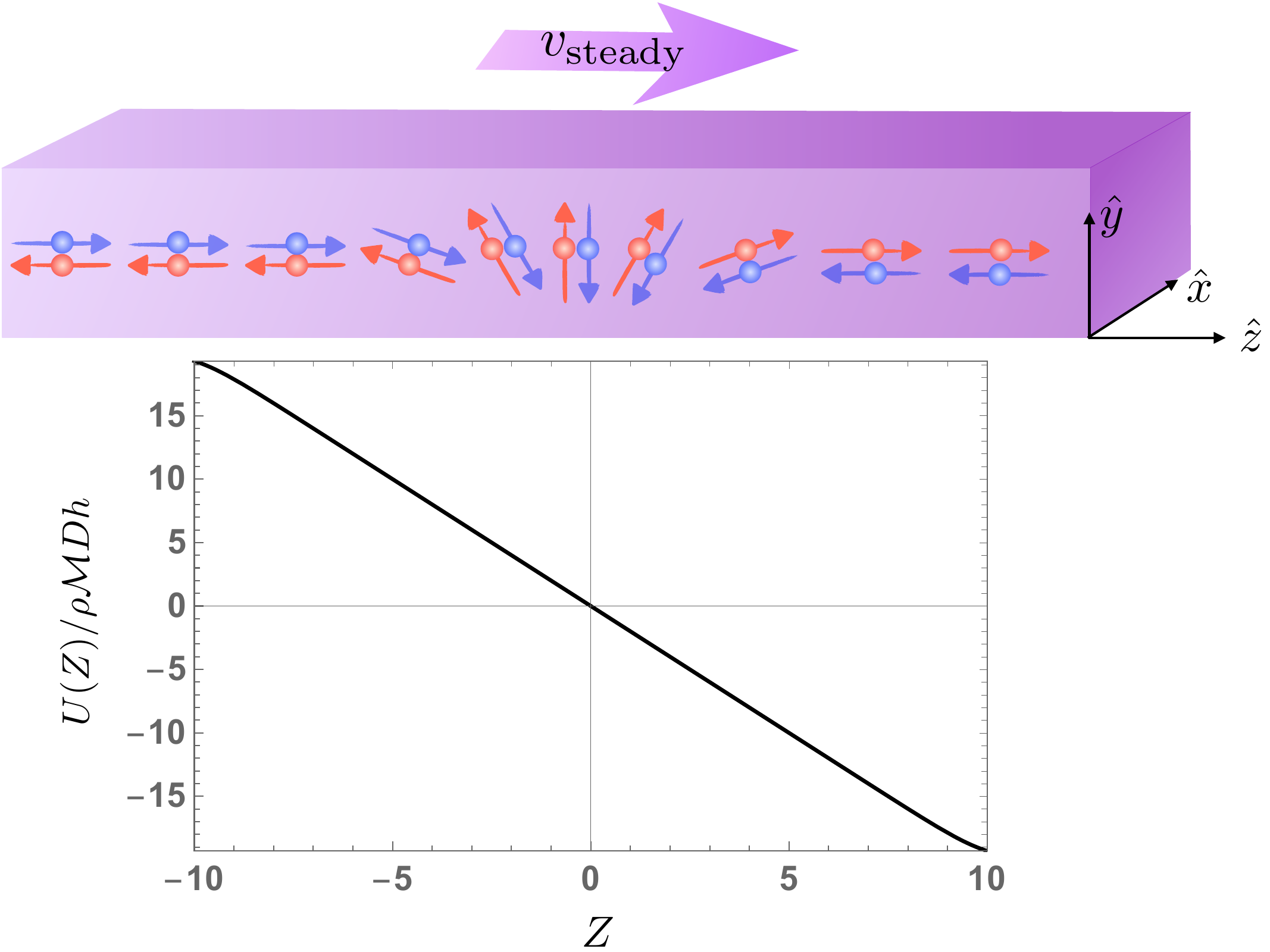}
\caption{A domain wall profile Eq.~\eqref{34} with $\Phi=\pi/2$ in a one-dimensional antiferromagnet. The graph shows the potential $U(Z)$, measured in unit of $\rho \mathcal{M}Dh\lambda$, for the domain wall in the presence of DM vector $\vb{D}=D\vb{\hat{y}}$ and magnetic field $\vb{h}=h\vb{\hat{x}}$. Here we set the size of the domain wall to be $\lambda=1$ and set the system to be $Z\in[-10,10]$. We see the potential is nearly linear and only bends close to two boundaries. The resultant constant forces acting on the domain wall, balanced with the dissipative force, leads to a  steady velocity $v_{\text{steady}}$ Eq.~\eqref{velocity}.}
\label{dwmotion}
\end{figure}

\begin{figure}
    \includegraphics[width=\columnwidth]{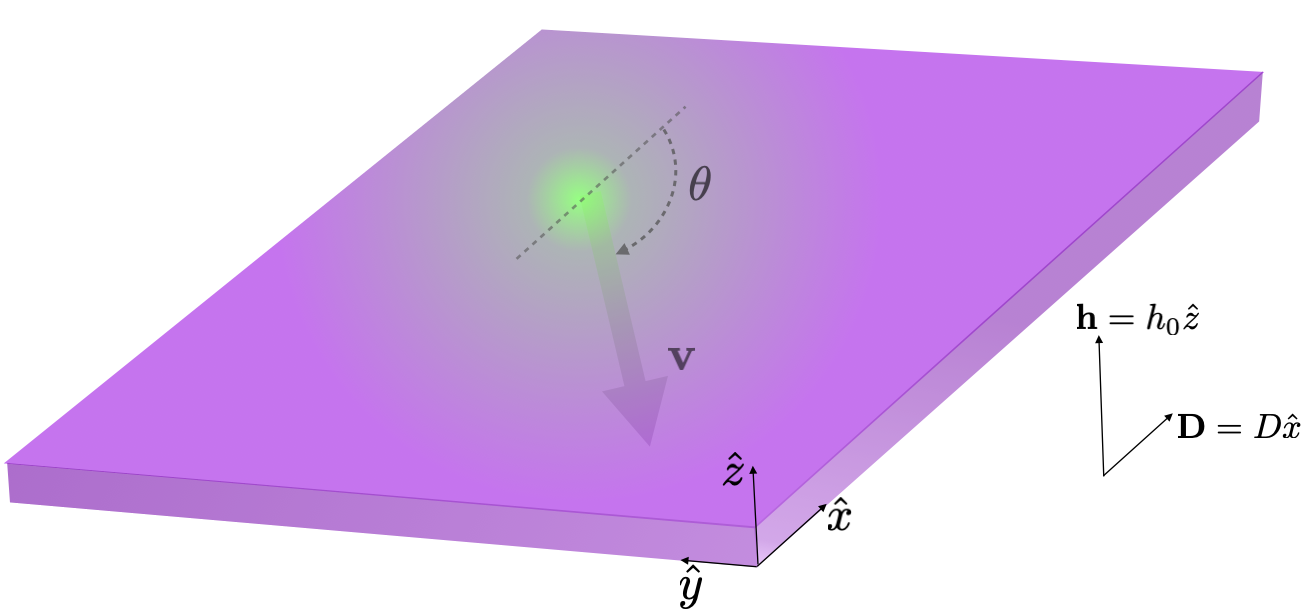}
\caption{The motion of an antiferromagnetic vortex core in the presence of DM vector $\mathbf{D}=D\mathbf{\hat{x}}$ and magnetic field $\mathbf{h}=h_0\mathbf{\hat{z}}$. The steady state velocity for the vortex is given in Eq.~\eqref{dhvortex} (setting $D_2=0$). The direction of the motion is determined by $\tan\theta=-M/gT$. }
\label{vortex}
\end{figure}

To cause a net displacement in the position of the soliton, we require: $(\mathbf{D}\times\mathbf{h})_{\bm{\hat{\zeta}}} \neq 0$, where $\bm{\hat{\zeta}}$ is the unit vector along a zero mode direction of the soliton. For example, for the domain wall $\bm{\hat{\zeta}}$ is along the easy-axis. This requires, in particular, a DM vector that is not aligned along the easy axis. To illustrate this idea, let us work out the dynamics of an easy-axis antiferromagnet in one dimension (see Fig.~\ref{dwmotion}), an antiferromagnetic vortex, with vorticity $\tilde{n} = 1$ in two dimensions (see Fig.~\ref{vortex}), and an antiferromagnetic hedgehog in three dimensions. 

\textit{Uniaxial domain wall:} We adopt the static domain walls parametrized in spherical angles $\theta(z)$ and $\phi(z)$ as follows
\(\cos\theta(z)=\tanh \frac{z-Z}{\lambda}, \;\; \phi(z)=\Phi,  \label{34} \)
where position $Z$ and azimuthal angle $\Phi$ are two collective coordinates, standing for two zero modes of the system. We now expand the first term in the Lagrangian Eq.~\eqref{eq.L-n-D-H}:
\[\mathcal{L}&=&\frac{\rho}{2}\dot{\vb{n}}^2 + \frac{\rho}{2} |\gamma \vb{n}\times \vb{h}|^2 +\frac{\rho}{2} \left(\frac{\mathcal{M}\vb{D}}{\gamma}\right)^2 \nonumber \\
 &+& \rho \gamma \dot{\vb{n}}\cdot (\vb{n}\times \vb{h}) - \frac{\rho \mathcal{M}}{\gamma}\dot{\vb{n}}\cdot\vb{D} \nonumber \\
  &-&\rho \mathcal{M}\vb{D}\cdot (\vb{n}\times \vb{h}). \label{35} \]
  We assume a simple configuration with $\vb{D}=D\mathbf{\hat{y}}$ and $\vb{h}=h\mathbf{\hat{x}}$. Both $D$ and $h$ are constants. Here $\rho \dot{\vb{n}}^2/2$ endows the domain wall with a mass $M$. As shown before, the magnetic field modifies the easy axis anisotropy. The term proportional to $\vb{D}^2$ is a constant and thus does not enter the equation of motion of the domain wall. 
  
  The total ``electromagnetic" force acting on the domain wall Eq.~\eqref{34} along $\hat{z}$ direction, derived from the vector potential in the second line of Eq.~\eqref{35}, vanishes in this situation. The last line in Lagrangian Eq.~\eqref{35} gives rise to a potential energy for the domain wall: 
  \(U(Z)\equiv \rho \mathcal{M}Dh\int_{-L}^Ldz\, n_z=\rho \mathcal{M}Dh \lambda \ln \frac{\cosh[(Z-L)/\lambda]}{\cosh[(Z+L)/\lambda]},\)
where we have parametrized the one dimensional antiferromagnet with $z\in[-L,L]$. One can therefore write down the equation of motion for the domain wall: 
\(  M\ddot{Z}=-M\dot{Z}/T+F,  \label{37}\)
where $F\equiv -dU/dZ$. We consider the situation that the domain wall is far away from two boundaries of the 1D antiferromagnet. The force due to the potential $U(Z)$ is a constant $F=2\rho \mathcal{M} D h$, independent of the position of the domain wall, in this scenario (see Fig.~\ref{dwmotion}). The domain wall mass $M$ is
\( M=\rho\int dz \, \left|\dv{\vb{n}}{Z}\right|^2 =\frac{2\rho}{\lambda}. \)
From Eq.~\eqref{37}, we can read off the velocity of steady motion:
\( v_{\text{steady}}=\mathcal{M}Dh T\lambda= 2\rho\mathcal{M} \frac{DhT}{M},  \label{velocity}\)
which is linearly proportional to the strength of DM interaction, applied magnetic field, viscous relaxation time, and is inversely proportional to the mass of domain wall, as one may expect. Note that the mass $M$ has a lower bound $\rho/L$, set by the system size. We also remark that one cannot crank up the magnetic field incontinently, as it also contributes to magnetization (recall $\vb{m}=\chi \vb{h}/\mathcal{M}+\cdots$, when $\vb{h}\perp \vb{n}$), which would ultimately invalidate our description at large fields. 

\textit{Antiferromagnetic vortex:} The dynamics of the antiferromagnetic vortex in the presence of an external out-of-plane magnetic field, $\mathbf{h} = h_0 \mathbf{\hat{z}}$, and an in-plane DM interaction mirrors the Magnus force dynamics presented in \cite{dasgupta17}. Magnus force type dynamics is unexpected and novel in the broader context of antiferromagnetic solitons with this as a possible new addition, for a review see \cite{Galkina2018}.

With an out of plane magnetic field a finite Skyrmion charge is generated for the antiferromagnetic vortex, $q = (\tilde{n}\rho\gamma h_0/2\mathcal{J})$ \cite{ivanov94,dasgupta17,Galkina2018}. Here $\tilde{n}$ is the winding number of the vortex. This in turn creates a finite gyromagnetic density $G_{XY} = - G_{YX} = g = 2\pi \tilde{n} \rho \gamma h_0$. This effect is notably absent for an in-plane magnetic field.

The in plane DM interaction provides a finite potential energy in the vortex center coordinate channels. With a DM vector of the form $\mathbf{D} = (D_1,D_2,0)$ we get a `Zeeman' energy density:
\begin{eqnarray}
\label{eq.vortex-D-H}
    U &=& \rho \mathcal{M} h_0 \int dxdy (-D_2 n_x + D_1 n_y) \\ \nonumber
    &=&  \rho \mathcal{M} h_0 \eta (-D_2 X + D_1 Y),
\end{eqnarray}
where $\eta$ is a structural factor that depends on the dimensions of the sample, see Fig.\ref{fig:potential-DM-h-AFM-vortex}. We provide an estimate for a sample with a circular geometry in the Appendix.\ref{sec.AFM-vortex}. This energy density is analogous to that of a planar ferromagnetic vortex with an in-plane magnetic field \cite{clarke08}. Assuming a circular geometry, $\eta = \pi R$, the force acting on our antiferromagnetic vortex is then $\mathbf{F} = -\pi R\rho \mathcal{M} h_0 (D_2, -D_1)$. Here R is the radius of the sample.

\begin{figure}[t]
\includegraphics[width=\columnwidth]{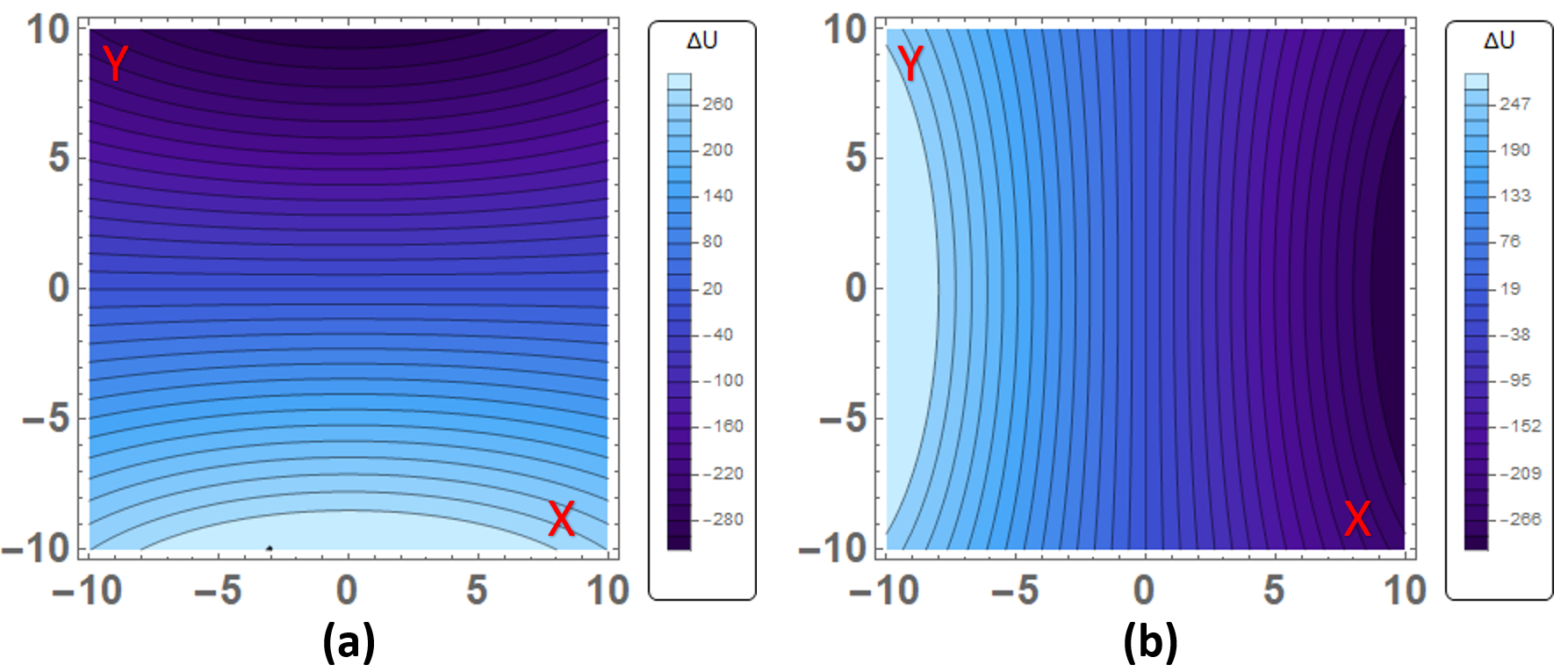}
\caption{Potential energy densities for the `Zeeman' $\mathcal{U} \propto \mathbf{n}\cdot(\mathbf{h}\times\mathbf{D})$ term in the collective coordinate space. The magnetic field is out of plane $\mathbf{h} = h_0\mathbf{\hat{z}}$ and the DM interaction is in plane. The sample geometry used was a square of side 20 units. On the left (a) is the density produced by $\mathbf{D} = (D_1,0)$ and on the right is the density produced by $\mathbf{D} = (0,D_2)$. As noted in the text the densities are linear in $X$ and $Y$ in the regime where the vortex core is well inside the boundary.}
\label{fig:potential-DM-h-AFM-vortex}
\end{figure}

The collective coordinate equations of motion for the vortex core reads:
\begin{eqnarray}
    M\ddot{X} &=& F_X + g \dot{Y} - \frac{M}{T} \dot{X}, \\ \nonumber
    M\ddot{Y} &=& F_Y - g \dot{X} - \frac{M}{T} \dot{Y}.
\end{eqnarray}
This gives the steady state velocity for the vortex core as:
\begin{equation}
    \mathbf{v} = T\rho\mathcal{M} h_0 R \pi\left(\frac{D_1 g T - D_2 M}{M^2 + g^2 T^2} ,\frac{D_2 g T + D_1 M}{M^2 + g^2 T^2} \right), \label{dhvortex}
\end{equation}
with  magnitude being
\begin{equation}
 |\mathbf{v}|= \frac{\pi R \rho \mathcal{M}}{\sqrt{1+g^2T^2/M^2}} \frac{|\mathbf{D}|h_0T}{M}.
\end{equation}
We note that, similar to the steady motion of a domain wall Eq.~\eqref{velocity}, $|\mathbf{v}|\propto |\mathbf{D}|\pi R h_0T/M$ when the gyromagnetic density is small $g\ll M/T$.

This situation is simplified for antiferromagnetic solitons where $\mathbf{n}(\mathbf{r})$ configuration covers the whole unit sphere---skyrmions and hedgehogs \cite{dasgupta17,Galkina2018}. For the Skyrmion with the same configuration of DM vectors and magnetic field we get a steady state velocity $\mathbf{v}_{\text{skyr}} = \pi R_{sk} \rho \mathcal{M} h_0(T/M)(-D_2, D_1)$ with $R_{sk}$ as the skyrmion radius. The dynamics here is also notably independent of system dimensions as skyrmions are local defects unlike vortices.

\textit{Antiferromagnetic hedgehog:} To illustrate  the cross term \eqref{eq.cross-dm-h} can be used to efficiently drive a hedgehog~\cite{pylypovskyi2012} in three dimensional antiferromagnets, we consider an isotropic hedgehog configuration
$\mathbf{n}(\mathbf{r})=\mathbf{n}_0(\mathbf{r}-\mathbf{R}_c)$
with $\mathbf{n}_0(\mathbf{r})=\mathbf{r}/|\mathbf{r}|$ and collective coordinates $\mathbf{R}_c$. The `Zeeman' energy potential for the hedgehog is given by 
\begin{equation}
 U(\mathbf{R}_c)=\rho \mathcal{M} \int d^3\mathbf{r}\, \mathbf{h}\cdot \mathbf{D}\times \mathbf{n}(\mathbf{r}), 
\end{equation}
where $\mathbf{h}$ and $\mathbf{D}$ are the magnetic field and the DM vector, respectively. 
This potential gives rise to a constant force
\begin{equation}
\mathbf{F}\equiv -\nabla_{\mathbf{R}_c} U = \frac{4\pi R^2}{3} \rho \mathcal{M}  \mathbf{h}\times \mathbf{D},  
\end{equation}
when the hedgehog is far away from the boundary of the sample. Note this force is dependent on the radius $R$ of the sample, which is a general feature of nonlocal solitons. The steady state velocity is 
\begin{equation}
\mathbf{v}= \frac{4\pi R^2}{3} \rho\mathcal{M} \frac{T\,\mathbf{h}\times \mathbf{D}}{M},
\end{equation}
whose direction is dictated by the orientations of the magnetic field and the DM vector. 

\subsection{DM interaction and adiabatic spin transfer torque}
A combination of these two interactions produces the Lagrangian density 
\begin{equation}
    \label{eq.n-u-d}
    \mathcal{L} = \frac{\rho (\dot{\mathbf{n}} + \mathbf{u}\cdot\bm{\nabla}\mathbf{n} + \mathcal{J}\mathbf{D})^2}{2}.
\end{equation}
The cross term generated here is proportional to $u_i(\partial_in_j)D_j$. This is clearly a total derivative term which has no effect in the bulk of a material where the DM vector is constant. However, at all interfaces and edges of the sample where the DM vector changes direction or magnitude or both, this term has a finite contribution. Across a sample boundary $\perp$'r to $x_i$, this term adds an energy:
\begin{equation}
    U_{\text{boundary}} = \rho\mathcal{J}u_i\mathbf{D}\cdot(\Delta_i\mathbf{n}).
\end{equation}
Depending on the sign of the DM interaction, the system will then choose to have the N{\'e}el vector along a boundary to orient $||$ to or $\perp$'r to the DM vector. Note that this boundary anisotropy is controlled by the direction of the adiabatic spin transfer torque $\mathbf{u}$, as the gradient is attached to that term.

\section{Discussion}
\label{sec.conc}
In this paper we studied the two sublattice antiferromagnet in the presence of external perturbations. The method we employed was to write  a field theory for the N{\'e}el field $\mathbf{n}$ and the uniform magnetization $\mathbf{m}$. The perturbations can then couple to these fields. One of our primary points is that to effectively move antiferromagnetic solitons we need to consider a combination of external perturbations. In all of this, our main motive is to identify avenues through which internal features like the inertia of $\mathbf{n}$, the location of solitons and their sizes can be controlled externally.

We work out these couplings for strain fields which modifies the inertia of the N{\'e}el vector. It does so by coupling through a magneto-elastic coupling. An external magnetic field can be put to multiple uses. A static field modifies the shape of the soliton and its configuration. It can also be used to create a local Berry curvature which can be coupled to using a spin current \cite{dasgupta17}. A dynamical magnetic field, $\mathbf{h}(\mathbf{r},t)$, can be used to generate an effective electromagnetic tensor and propel domain walls \cite{yamane17}. Crucially, what we find, is that in combination with a Dzyaloshinsky-Moriya interaction a magnetic field acts to provide a Zeeman like interaction for the N{\'e}el vector which can be used to directly drive the antiferromagnetic soliton.

\section*{Acknowledgements}
The authors would like to thank Oleg Tchernyshyov,  Se Kwon Kim and Yaroslav Tserkovnyak for their useful comments. This work was supported by NSERC, the Max Planck-UBC-UTokyo Centre for Quantum Materials and  the  Canada First Research Excellence Fund, Quantum Materials and Future Technologies Program and the Japan Society for the Promotion of Science KAKENHI Grant No. JP19H01808. S.D. was supported though the MPI-UBC-UTokyo postdoctoral fellowship. J.Z. was supported by NSF under Grant No. DMR-1742928.

\appendix
\counterwithin{figure}{section}{equation}
\section{Energy density of the antiferromagnetic vortex}
\label{sec.AFM-vortex}
Here we provide the details for the energy density of the antiferromagnetic vortex in the presence of an out of plane magnetic field and an in-plane DM interactions. The configuration we use is $\mathbf{D} = (D_1,D_2,0)$ and $h = h_0\mathbf{\hat{z}}$. The energy density is given by the first line of Eq.~(\ref{eq.vortex-D-H}).

To calculate a form for the energy density let us consider a vortex profile of the form, $\mathbf{n} = \frac{(x,~y,~0)}{\sqrt{x^2 + y^2}}$ with the core at the origin. To calculate the energy density what we do is displace the vortex core slightly from the origin $(0,0) \to (\delta_x,\delta_y)$ and subtract the two energies of the two spin profiles.
\begin{equation}
    \Delta \mathcal{U} = \mathcal{U}[X = \delta_x, Y = \delta_y] - \mathcal{U}[X = 0, Y = 0],
\end{equation}
which is then integrated over space $\int dx dy~ \Delta \mathcal{U}$ to get the energy in the collective coordinate space $\Delta U$. The force is then---$\mathbf{F} = -(\Delta U/\delta_x,\Delta U/\delta_x)$. The displaced energy density is then:
\begin{eqnarray}
     U &=& \rho \mathcal{M} h_0 \int dxdy~ (-D_2 n_x + D_1 n_y) \\ \nonumber
     \Delta U &=& \rho \mathcal{M} h_0 \left[ D_{2}(I_1 \delta_x - I_2 \delta_y) +  D_1 ( I_2 \delta_x - I_3\delta_y) \right], 
\end{eqnarray}
with the integrals
\begin{eqnarray} \nonumber
     I_1 &=& \int dx dy \frac{y^2}{(x^2 + y^2)^{3/2}},~ I_3 = \int dx dy \frac{x^2}{(x^2 + y^2)^{3/2}} \\ 
     I_2 &=& \int dx dy \frac{xy}{(x^2 + y^2)^{3/2}}. 
\end{eqnarray}
Under the assumption that we are working with a symmetric sample we can see that $I_2 = 0$. The other two integrals, $I_1$ and $I_3$ need to be worked out for specific sample geometries. We can analytically work out the very simple situation where the sample has a circular geometry of radius $R$ and the vortex core is displaced slightly. In this case we can convert the integrals to spherical coordinates on the plane and we obtain $I_1 = I_3 = \pi R$. This then produces the energy $\Delta U = \rho \mathcal{M}\pi R h_0 (D_2 \delta_x - D_1\delta_y)$ and the force $\mathbf{F} = \rho \mathcal{M}\pi R h_0 (-D_2, D_1)$.

\bibliographystyle{apsrev4-1}
\bibliography{main}

\begin{thebibliography}{45}%
\makeatletter
\providecommand \@ifxundefined [1]{%
 \@ifx{#1\undefined}
}%
\providecommand \@ifnum [1]{%
 \ifnum #1\expandafter \@firstoftwo
 \else \expandafter \@secondoftwo
 \fi
}%
\providecommand \@ifx [1]{%
 \ifx #1\expandafter \@firstoftwo
 \else \expandafter \@secondoftwo
 \fi
}%
\providecommand \natexlab [1]{#1}%
\providecommand \enquote  [1]{``#1''}%
\providecommand \bibnamefont  [1]{#1}%
\providecommand \bibfnamefont [1]{#1}%
\providecommand \citenamefont [1]{#1}%
\providecommand \href@noop [0]{\@secondoftwo}%
\providecommand \href [0]{\begingroup \@sanitize@url \@href}%
\providecommand \@href[1]{\@@startlink{#1}\@@href}%
\providecommand \@@href[1]{\endgroup#1\@@endlink}%
\providecommand \@sanitize@url [0]{\catcode `\\12\catcode `\$12\catcode
  `\&12\catcode `\#12\catcode `\^12\catcode `\_12\catcode `\%12\relax}%
\providecommand \@@startlink[1]{}%
\providecommand \@@endlink[0]{}%
\providecommand \url  [0]{\begingroup\@sanitize@url \@url }%
\providecommand \@url [1]{\endgroup\@href {#1}{\urlprefix }}%
\providecommand \urlprefix  [0]{URL }%
\providecommand \Eprint [0]{\href }%
\providecommand \doibase [0]{http://dx.doi.org/}%
\providecommand \selectlanguage [0]{\@gobble}%
\providecommand \bibinfo  [0]{\@secondoftwo}%
\providecommand \bibfield  [0]{\@secondoftwo}%
\providecommand \translation [1]{[#1]}%
\providecommand \BibitemOpen [0]{}%
\providecommand \bibitemStop [0]{}%
\providecommand \bibitemNoStop [0]{.\EOS\space}%
\providecommand \EOS [0]{\spacefactor3000\relax}%
\providecommand \BibitemShut  [1]{\csname bibitem#1\endcsname}%
\let\auto@bib@innerbib\@empty
\bibitem [{\citenamefont {N\'u\~nez}\ \emph {et~al.}(2006)\citenamefont
  {N\'u\~nez}, \citenamefont {Duine}, \citenamefont {Haney},\ and\
  \citenamefont {MacDonald}}]{nunez2006}%
  \BibitemOpen
  \bibfield  {author} {\bibinfo {author} {\bibfnamefont {A.~S.}\ \bibnamefont
  {N\'u\~nez}}, \bibinfo {author} {\bibfnamefont {R.~A.}\ \bibnamefont
  {Duine}}, \bibinfo {author} {\bibfnamefont {P.}~\bibnamefont {Haney}}, \ and\
  \bibinfo {author} {\bibfnamefont {A.~H.}\ \bibnamefont {MacDonald}},\ }\href
  {\doibase 10.1103/PhysRevB.73.214426} {\bibfield  {journal} {\bibinfo
  {journal} {Phys. Rev. B}\ }\textbf {\bibinfo {volume} {73}},\ \bibinfo
  {pages} {214426} (\bibinfo {year} {2006})}\BibitemShut {NoStop}%
\bibitem [{\citenamefont {Baltz}\ \emph {et~al.}(2018)\citenamefont {Baltz},
  \citenamefont {Manchon}, \citenamefont {Tsoi}, \citenamefont {Moriyama},
  \citenamefont {Ono},\ and\ \citenamefont {Tserkovnyak}}]{RMPYaroslav2018}%
  \BibitemOpen
  \bibfield  {author} {\bibinfo {author} {\bibfnamefont {V.}~\bibnamefont
  {Baltz}}, \bibinfo {author} {\bibfnamefont {A.}~\bibnamefont {Manchon}},
  \bibinfo {author} {\bibfnamefont {M.}~\bibnamefont {Tsoi}}, \bibinfo {author}
  {\bibfnamefont {T.}~\bibnamefont {Moriyama}}, \bibinfo {author}
  {\bibfnamefont {T.}~\bibnamefont {Ono}}, \ and\ \bibinfo {author}
  {\bibfnamefont {Y.}~\bibnamefont {Tserkovnyak}},\ }\href {\doibase
  10.1103/RevModPhys.90.015005} {\bibfield  {journal} {\bibinfo  {journal}
  {Rev. Mod. Phys.}\ }\textbf {\bibinfo {volume} {90}},\ \bibinfo {pages}
  {015005} (\bibinfo {year} {2018})}\BibitemShut {NoStop}%
\bibitem [{\citenamefont {Shick}\ \emph {et~al.}(2010)\citenamefont {Shick},
  \citenamefont {Khmelevskyi}, \citenamefont {Mryasov}, \citenamefont
  {Wunderlich},\ and\ \citenamefont {Jungwirth}}]{shick2010}%
  \BibitemOpen
  \bibfield  {author} {\bibinfo {author} {\bibfnamefont {A.~B.}\ \bibnamefont
  {Shick}}, \bibinfo {author} {\bibfnamefont {S.}~\bibnamefont {Khmelevskyi}},
  \bibinfo {author} {\bibfnamefont {O.~N.}\ \bibnamefont {Mryasov}}, \bibinfo
  {author} {\bibfnamefont {J.}~\bibnamefont {Wunderlich}}, \ and\ \bibinfo
  {author} {\bibfnamefont {T.}~\bibnamefont {Jungwirth}},\ }\href {\doibase
  10.1103/PhysRevB.81.212409} {\bibfield  {journal} {\bibinfo  {journal} {Phys.
  Rev. B}\ }\textbf {\bibinfo {volume} {81}},\ \bibinfo {pages} {212409}
  (\bibinfo {year} {2010})}\BibitemShut {NoStop}%
\bibitem [{\citenamefont {Jungwirth}\ \emph {et~al.}(2016)\citenamefont
  {Jungwirth}, \citenamefont {Marti}, \citenamefont {Wadley},\ and\
  \citenamefont {Wunderlich}}]{Jungwirth2016}%
  \BibitemOpen
  \bibfield  {author} {\bibinfo {author} {\bibfnamefont {T.}~\bibnamefont
  {Jungwirth}}, \bibinfo {author} {\bibfnamefont {X.}~\bibnamefont {Marti}},
  \bibinfo {author} {\bibfnamefont {P.}~\bibnamefont {Wadley}}, \ and\ \bibinfo
  {author} {\bibfnamefont {J.}~\bibnamefont {Wunderlich}},\ }\href {\doibase
  10.1038/nnano.2016.18} {\bibfield  {journal} {\bibinfo  {journal} {Nature
  Nanotechnology}\ }\textbf {\bibinfo {volume} {11}},\ \bibinfo {pages} {231}
  (\bibinfo {year} {2016})}\BibitemShut {NoStop}%
\bibitem [{\citenamefont {Marrows}(2016)}]{Marrows2016}%
  \BibitemOpen
  \bibfield  {author} {\bibinfo {author} {\bibfnamefont {C.}~\bibnamefont
  {Marrows}},\ }\href {\doibase 10.1126/science.aad8211} {\bibfield  {journal}
  {\bibinfo  {journal} {Science}\ }\textbf {\bibinfo {volume} {351}},\ \bibinfo
  {pages} {558} (\bibinfo {year} {2016})}\BibitemShut {NoStop}%
\bibitem [{\citenamefont {Kim}\ \emph {et~al.}(2014)\citenamefont {Kim},
  \citenamefont {Tserkovnyak},\ and\ \citenamefont {Tchernyshyov}}]{skkim14}%
  \BibitemOpen
  \bibfield  {author} {\bibinfo {author} {\bibfnamefont {S.~K.}\ \bibnamefont
  {Kim}}, \bibinfo {author} {\bibfnamefont {Y.}~\bibnamefont {Tserkovnyak}}, \
  and\ \bibinfo {author} {\bibfnamefont {O.}~\bibnamefont {Tchernyshyov}},\
  }\href {\doibase 10.1103/PhysRevB.90.104406} {\bibfield  {journal} {\bibinfo
  {journal} {Phys. Rev. B}\ }\textbf {\bibinfo {volume} {90}},\ \bibinfo
  {pages} {104406} (\bibinfo {year} {2014})}\BibitemShut {NoStop}%
\bibitem [{\citenamefont {Tveten}\ \emph {et~al.}(2014)\citenamefont {Tveten},
  \citenamefont {Qaiumzadeh},\ and\ \citenamefont {Brataas}}]{tveten14}%
  \BibitemOpen
  \bibfield  {author} {\bibinfo {author} {\bibfnamefont {E.~G.}\ \bibnamefont
  {Tveten}}, \bibinfo {author} {\bibfnamefont {A.}~\bibnamefont {Qaiumzadeh}},
  \ and\ \bibinfo {author} {\bibfnamefont {A.}~\bibnamefont {Brataas}},\ }\href
  {\doibase 10.1103/PhysRevLett.112.147204} {\bibfield  {journal} {\bibinfo
  {journal} {Phys. Rev. Lett.}\ }\textbf {\bibinfo {volume} {112}},\ \bibinfo
  {pages} {147204} (\bibinfo {year} {2014})}\BibitemShut {NoStop}%
\bibitem [{\citenamefont {Qaiumzadeh}\ \emph {et~al.}(2018)\citenamefont
  {Qaiumzadeh}, \citenamefont {Kristiansen},\ and\ \citenamefont
  {Brataas}}]{qaiumzadeh18}%
  \BibitemOpen
  \bibfield  {author} {\bibinfo {author} {\bibfnamefont {A.}~\bibnamefont
  {Qaiumzadeh}}, \bibinfo {author} {\bibfnamefont {L.~A.}\ \bibnamefont
  {Kristiansen}}, \ and\ \bibinfo {author} {\bibfnamefont {A.}~\bibnamefont
  {Brataas}},\ }\href {\doibase 10.1103/PhysRevB.97.020402} {\bibfield
  {journal} {\bibinfo  {journal} {Phys. Rev. B}\ }\textbf {\bibinfo {volume}
  {97}},\ \bibinfo {pages} {020402} (\bibinfo {year} {2018})}\BibitemShut
  {NoStop}%
\bibitem [{\citenamefont {Dasgupta}\ \emph {et~al.}(2017)\citenamefont
  {Dasgupta}, \citenamefont {Kim},\ and\ \citenamefont
  {Tchernyshyov}}]{dasgupta17}%
  \BibitemOpen
  \bibfield  {author} {\bibinfo {author} {\bibfnamefont {S.}~\bibnamefont
  {Dasgupta}}, \bibinfo {author} {\bibfnamefont {S.~K.}\ \bibnamefont {Kim}}, \
  and\ \bibinfo {author} {\bibfnamefont {O.}~\bibnamefont {Tchernyshyov}},\
  }\href {\doibase 10.1103/PhysRevB.95.220407} {\bibfield  {journal} {\bibinfo
  {journal} {Phys. Rev. B}\ }\textbf {\bibinfo {volume} {95}},\ \bibinfo
  {pages} {220407} (\bibinfo {year} {2017})}\BibitemShut {NoStop}%
\bibitem [{\citenamefont {Schryer}\ and\ \citenamefont
  {Walker}(1974)}]{schryer74}%
  \BibitemOpen
  \bibfield  {author} {\bibinfo {author} {\bibfnamefont {N.~L.}\ \bibnamefont
  {Schryer}}\ and\ \bibinfo {author} {\bibfnamefont {L.~R.}\ \bibnamefont
  {Walker}},\ }\href {\doibase 10.1063/1.1663252} {\bibfield  {journal}
  {\bibinfo  {journal} {Journal of Applied Physics}\ }\textbf {\bibinfo
  {volume} {45}},\ \bibinfo {pages} {5406} (\bibinfo {year}
  {1974})}\BibitemShut {NoStop}%
\bibitem [{\citenamefont {Tretiakov}\ \emph {et~al.}(2008)\citenamefont
  {Tretiakov}, \citenamefont {Clarke}, \citenamefont {Chern}, \citenamefont
  {Bazaliy},\ and\ \citenamefont {Tchernyshyov}}]{tretiakov08}%
  \BibitemOpen
  \bibfield  {author} {\bibinfo {author} {\bibfnamefont {O.~A.}\ \bibnamefont
  {Tretiakov}}, \bibinfo {author} {\bibfnamefont {D.}~\bibnamefont {Clarke}},
  \bibinfo {author} {\bibfnamefont {G.-W.}\ \bibnamefont {Chern}}, \bibinfo
  {author} {\bibfnamefont {Y.~B.}\ \bibnamefont {Bazaliy}}, \ and\ \bibinfo
  {author} {\bibfnamefont {O.}~\bibnamefont {Tchernyshyov}},\ }\href {\doibase
  10.1103/PhysRevLett.100.127204} {\bibfield  {journal} {\bibinfo  {journal}
  {Phys. Rev. Lett.}\ }\textbf {\bibinfo {volume} {100}},\ \bibinfo {pages}
  {127204} (\bibinfo {year} {2008})}\BibitemShut {NoStop}%
\bibitem [{\citenamefont {Dasgupta}\ and\ \citenamefont
  {Tchernyshyov}(2020)}]{dasgupta2020}%
  \BibitemOpen
  \bibfield  {author} {\bibinfo {author} {\bibfnamefont {S.}~\bibnamefont
  {Dasgupta}}\ and\ \bibinfo {author} {\bibfnamefont {O.}~\bibnamefont
  {Tchernyshyov}},\ }\href {\doibase 10.1103/PhysRevB.102.144417} {\bibfield
  {journal} {\bibinfo  {journal} {Phys. Rev. B}\ }\textbf {\bibinfo {volume}
  {102}},\ \bibinfo {pages} {144417} (\bibinfo {year} {2020})}\BibitemShut
  {NoStop}%
\bibitem [{\citenamefont {Gomonay}\ \emph {et~al.}(2016)\citenamefont
  {Gomonay}, \citenamefont {Jungwirth},\ and\ \citenamefont
  {Sinova}}]{gomonay16}%
  \BibitemOpen
  \bibfield  {author} {\bibinfo {author} {\bibfnamefont {O.}~\bibnamefont
  {Gomonay}}, \bibinfo {author} {\bibfnamefont {T.}~\bibnamefont {Jungwirth}},
  \ and\ \bibinfo {author} {\bibfnamefont {J.}~\bibnamefont {Sinova}},\ }\href
  {\doibase 10.1103/PhysRevLett.117.017202} {\bibfield  {journal} {\bibinfo
  {journal} {Phys. Rev. Lett.}\ }\textbf {\bibinfo {volume} {117}},\ \bibinfo
  {pages} {017202} (\bibinfo {year} {2016})}\BibitemShut {NoStop}%
\bibitem [{\citenamefont {\ifmmode~\check{Z}\else \v{Z}\fi{}elezn\'y}\ \emph
  {et~al.}(2014)\citenamefont {\ifmmode~\check{Z}\else \v{Z}\fi{}elezn\'y},
  \citenamefont {Gao}, \citenamefont {V\'yborn\'y}, \citenamefont {Zemen},
  \citenamefont {Ma\ifmmode~\check{s}\else \v{s}\fi{}ek}, \citenamefont
  {Manchon}, \citenamefont {Wunderlich}, \citenamefont {Sinova},\ and\
  \citenamefont {Jungwirth}}]{zelezny14}%
  \BibitemOpen
  \bibfield  {author} {\bibinfo {author} {\bibfnamefont {J.}~\bibnamefont
  {\ifmmode~\check{Z}\else \v{Z}\fi{}elezn\'y}}, \bibinfo {author}
  {\bibfnamefont {H.}~\bibnamefont {Gao}}, \bibinfo {author} {\bibfnamefont
  {K.}~\bibnamefont {V\'yborn\'y}}, \bibinfo {author} {\bibfnamefont
  {J.}~\bibnamefont {Zemen}}, \bibinfo {author} {\bibfnamefont
  {J.}~\bibnamefont {Ma\ifmmode~\check{s}\else \v{s}\fi{}ek}}, \bibinfo
  {author} {\bibfnamefont {A.}~\bibnamefont {Manchon}}, \bibinfo {author}
  {\bibfnamefont {J.}~\bibnamefont {Wunderlich}}, \bibinfo {author}
  {\bibfnamefont {J.}~\bibnamefont {Sinova}}, \ and\ \bibinfo {author}
  {\bibfnamefont {T.}~\bibnamefont {Jungwirth}},\ }\href {\doibase
  10.1103/PhysRevLett.113.157201} {\bibfield  {journal} {\bibinfo  {journal}
  {Phys. Rev. Lett.}\ }\textbf {\bibinfo {volume} {113}},\ \bibinfo {pages}
  {157201} (\bibinfo {year} {2014})}\BibitemShut {NoStop}%
\bibitem [{\citenamefont {Wadley}\ \emph {et~al.}(2016)\citenamefont {Wadley},
  \citenamefont {Howells}, \citenamefont {{\v{Z}}elezn{\'y}}, \citenamefont
  {Andrews}, \citenamefont {Hills}, \citenamefont {Campion}, \citenamefont
  {Nov{\'a}k}, \citenamefont {Olejn{\'i}k}, \citenamefont {Maccherozzi},
  \citenamefont {Dhesi}, \citenamefont {Martin}, \citenamefont {Wagner},
  \citenamefont {Wunderlich}, \citenamefont {Freimuth}, \citenamefont
  {Mokrousov}, \citenamefont {Kune{\v{s}}}, \citenamefont {Chauhan},
  \citenamefont {Grzybowski}, \citenamefont {Rushforth}, \citenamefont
  {Edmonds}, \citenamefont {Gallagher},\ and\ \citenamefont
  {Jungwirth}}]{Wadley2016}%
  \BibitemOpen
  \bibfield  {author} {\bibinfo {author} {\bibfnamefont {P.}~\bibnamefont
  {Wadley}}, \bibinfo {author} {\bibfnamefont {B.}~\bibnamefont {Howells}},
  \bibinfo {author} {\bibfnamefont {J.}~\bibnamefont {{\v{Z}}elezn{\'y}}},
  \bibinfo {author} {\bibfnamefont {C.}~\bibnamefont {Andrews}}, \bibinfo
  {author} {\bibfnamefont {V.}~\bibnamefont {Hills}}, \bibinfo {author}
  {\bibfnamefont {R.~P.}\ \bibnamefont {Campion}}, \bibinfo {author}
  {\bibfnamefont {V.}~\bibnamefont {Nov{\'a}k}}, \bibinfo {author}
  {\bibfnamefont {K.}~\bibnamefont {Olejn{\'i}k}}, \bibinfo {author}
  {\bibfnamefont {F.}~\bibnamefont {Maccherozzi}}, \bibinfo {author}
  {\bibfnamefont {S.~S.}\ \bibnamefont {Dhesi}}, \bibinfo {author}
  {\bibfnamefont {S.~Y.}\ \bibnamefont {Martin}}, \bibinfo {author}
  {\bibfnamefont {T.}~\bibnamefont {Wagner}}, \bibinfo {author} {\bibfnamefont
  {J.}~\bibnamefont {Wunderlich}}, \bibinfo {author} {\bibfnamefont
  {F.}~\bibnamefont {Freimuth}}, \bibinfo {author} {\bibfnamefont
  {Y.}~\bibnamefont {Mokrousov}}, \bibinfo {author} {\bibfnamefont
  {J.}~\bibnamefont {Kune{\v{s}}}}, \bibinfo {author} {\bibfnamefont {J.~S.}\
  \bibnamefont {Chauhan}}, \bibinfo {author} {\bibfnamefont {M.~J.}\
  \bibnamefont {Grzybowski}}, \bibinfo {author} {\bibfnamefont {A.~W.}\
  \bibnamefont {Rushforth}}, \bibinfo {author} {\bibfnamefont {K.~W.}\
  \bibnamefont {Edmonds}}, \bibinfo {author} {\bibfnamefont {B.~L.}\
  \bibnamefont {Gallagher}}, \ and\ \bibinfo {author} {\bibfnamefont
  {T.}~\bibnamefont {Jungwirth}},\ }\href {\doibase 10.1126/science.aab1031}
  {\bibfield  {journal} {\bibinfo  {journal} {Science}\ }\textbf {\bibinfo
  {volume} {351}},\ \bibinfo {pages} {587} (\bibinfo {year}
  {2016})}\BibitemShut {NoStop}%
\bibitem [{\citenamefont {Dzyaloshinsky}(1958)}]{dzyaloshinsky58}%
  \BibitemOpen
  \bibfield  {author} {\bibinfo {author} {\bibfnamefont {I.}~\bibnamefont
  {Dzyaloshinsky}},\ }\href {\doibase
  https://doi.org/10.1016/0022-3697(58)90076-3} {\bibfield  {journal} {\bibinfo
   {journal} {J. of Phys. Chem. Sol.}\ }\textbf {\bibinfo {volume} {4}},\
  \bibinfo {pages} {241 } (\bibinfo {year} {1958})}\BibitemShut {NoStop}%
\bibitem [{\citenamefont {Moriya}(1960)}]{moriya60}%
  \BibitemOpen
  \bibfield  {author} {\bibinfo {author} {\bibfnamefont {T.}~\bibnamefont
  {Moriya}},\ }\href {\doibase 10.1103/PhysRev.120.91} {\bibfield  {journal}
  {\bibinfo  {journal} {Phys. Rev.}\ }\textbf {\bibinfo {volume} {120}},\
  \bibinfo {pages} {91} (\bibinfo {year} {1960})}\BibitemShut {NoStop}%
\bibitem [{\citenamefont {Thiele}(1973)}]{thiele73}%
  \BibitemOpen
  \bibfield  {author} {\bibinfo {author} {\bibfnamefont {A.~A.}\ \bibnamefont
  {Thiele}},\ }\href {\doibase 10.1103/PhysRevLett.30.230} {\bibfield
  {journal} {\bibinfo  {journal} {Phys. Rev. Lett.}\ }\textbf {\bibinfo
  {volume} {30}},\ \bibinfo {pages} {230} (\bibinfo {year} {1973})}\BibitemShut
  {NoStop}%
\bibitem [{\citenamefont {Bazaliy}\ \emph {et~al.}(1998)\citenamefont
  {Bazaliy}, \citenamefont {Jones},\ and\ \citenamefont {Zhang}}]{bazaliy98}%
  \BibitemOpen
  \bibfield  {author} {\bibinfo {author} {\bibfnamefont {Y.~B.}\ \bibnamefont
  {Bazaliy}}, \bibinfo {author} {\bibfnamefont {B.~A.}\ \bibnamefont {Jones}},
  \ and\ \bibinfo {author} {\bibfnamefont {S.-C.}\ \bibnamefont {Zhang}},\
  }\href {\doibase 10.1103/PhysRevB.57.R3213} {\bibfield  {journal} {\bibinfo
  {journal} {Phys. Rev. B}\ }\textbf {\bibinfo {volume} {57}},\ \bibinfo
  {pages} {R3213} (\bibinfo {year} {1998})}\BibitemShut {NoStop}%
\bibitem [{\citenamefont {Slonczewski}(2002)}]{slonczewski02}%
  \BibitemOpen
  \bibfield  {author} {\bibinfo {author} {\bibfnamefont {J.}~\bibnamefont
  {Slonczewski}},\ }\href {\doibase
  https://doi.org/10.1016/S0304-8853(02)00291-3} {\bibfield  {journal}
  {\bibinfo  {journal} {Journal of Magnetism and Magnetic Materials}\ }\textbf
  {\bibinfo {volume} {247}},\ \bibinfo {pages} {324} (\bibinfo {year}
  {2002})}\BibitemShut {NoStop}%
\bibitem [{\citenamefont {Haldane}(1986)}]{haldane86}%
  \BibitemOpen
  \bibfield  {author} {\bibinfo {author} {\bibfnamefont {F.~D.~M.}\
  \bibnamefont {Haldane}},\ }\href {\doibase 10.1103/PhysRevLett.57.1488}
  {\bibfield  {journal} {\bibinfo  {journal} {Phys. Rev. Lett.}\ }\textbf
  {\bibinfo {volume} {57}},\ \bibinfo {pages} {1488} (\bibinfo {year}
  {1986})}\BibitemShut {NoStop}%
\bibitem [{\citenamefont {Ivanov}\ and\ \citenamefont
  {Kolezhuk}(1995)}]{ivanov95}%
  \BibitemOpen
  \bibfield  {author} {\bibinfo {author} {\bibfnamefont {B.~A.}\ \bibnamefont
  {Ivanov}}\ and\ \bibinfo {author} {\bibfnamefont {A.~K.}\ \bibnamefont
  {Kolezhuk}},\ }\href {\doibase 10.1103/PhysRevLett.74.1859} {\bibfield
  {journal} {\bibinfo  {journal} {Phys. Rev. Lett.}\ }\textbf {\bibinfo
  {volume} {74}},\ \bibinfo {pages} {1859} (\bibinfo {year}
  {1995})}\BibitemShut {NoStop}%
\bibitem [{\citenamefont {Dasgupta}\ and\ \citenamefont
  {Tchernyshyov}(2018)}]{dasgupta18}%
  \BibitemOpen
  \bibfield  {author} {\bibinfo {author} {\bibfnamefont {S.}~\bibnamefont
  {Dasgupta}}\ and\ \bibinfo {author} {\bibfnamefont {O.}~\bibnamefont
  {Tchernyshyov}},\ }\href {\doibase 10.1103/PhysRevB.98.224401} {\bibfield
  {journal} {\bibinfo  {journal} {Phys. Rev. B}\ }\textbf {\bibinfo {volume}
  {98}},\ \bibinfo {pages} {224401} (\bibinfo {year} {2018})}\BibitemShut
  {NoStop}%
\bibitem [{\citenamefont {Dasgupta}\ \emph {et~al.}(2020)\citenamefont
  {Dasgupta}, \citenamefont {Zhang}, \citenamefont {Bah},\ and\ \citenamefont
  {Tchernyshyov}}]{dasgupta20}%
  \BibitemOpen
  \bibfield  {author} {\bibinfo {author} {\bibfnamefont {S.}~\bibnamefont
  {Dasgupta}}, \bibinfo {author} {\bibfnamefont {S.}~\bibnamefont {Zhang}},
  \bibinfo {author} {\bibfnamefont {I.}~\bibnamefont {Bah}}, \ and\ \bibinfo
  {author} {\bibfnamefont {O.}~\bibnamefont {Tchernyshyov}},\ }\href {\doibase
  10.1103/PhysRevLett.124.157203} {\bibfield  {journal} {\bibinfo  {journal}
  {Phys. Rev. Lett.}\ }\textbf {\bibinfo {volume} {124}},\ \bibinfo {pages}
  {157203} (\bibinfo {year} {2020})}\BibitemShut {NoStop}%
\bibitem [{\citenamefont {Tchernyshyov}\ \emph {et~al.}(2002)\citenamefont
  {Tchernyshyov}, \citenamefont {Moessner},\ and\ \citenamefont
  {Sondhi}}]{tchernyshyov02}%
  \BibitemOpen
  \bibfield  {author} {\bibinfo {author} {\bibfnamefont {O.}~\bibnamefont
  {Tchernyshyov}}, \bibinfo {author} {\bibfnamefont {R.}~\bibnamefont
  {Moessner}}, \ and\ \bibinfo {author} {\bibfnamefont {S.~L.}\ \bibnamefont
  {Sondhi}},\ }\href {\doibase 10.1103/PhysRevB.66.064403} {\bibfield
  {journal} {\bibinfo  {journal} {Phys. Rev. B}\ }\textbf {\bibinfo {volume}
  {66}},\ \bibinfo {pages} {064403} (\bibinfo {year} {2002})}\BibitemShut
  {NoStop}%
\bibitem [{\citenamefont {Cable}\ \emph {et~al.}(1993)\citenamefont {Cable},
  \citenamefont {Wakabayashi},\ and\ \citenamefont {Radhakrishna}}]{cable1993}%
  \BibitemOpen
  \bibfield  {author} {\bibinfo {author} {\bibfnamefont {J.~W.}\ \bibnamefont
  {Cable}}, \bibinfo {author} {\bibfnamefont {N.}~\bibnamefont {Wakabayashi}},
  \ and\ \bibinfo {author} {\bibfnamefont {P.}~\bibnamefont {Radhakrishna}},\
  }\href {\doibase 10.1103/PhysRevB.48.6159} {\bibfield  {journal} {\bibinfo
  {journal} {Phys. Rev. B}\ }\textbf {\bibinfo {volume} {48}},\ \bibinfo
  {pages} {6159} (\bibinfo {year} {1993})}\BibitemShut {NoStop}%
\bibitem [{\citenamefont {Chen}\ \emph {et~al.}(2020)\citenamefont {Chen},
  \citenamefont {Gaudet}, \citenamefont {Dasgupta}, \citenamefont {Marcus},
  \citenamefont {Lin}, \citenamefont {Chen}, \citenamefont {Tomita},
  \citenamefont {Ikhlas}, \citenamefont {Zhao}, \citenamefont {Chen},
  \citenamefont {Stone}, \citenamefont {Tchernyshyov}, \citenamefont
  {Nakatsuji},\ and\ \citenamefont {Broholm}}]{chen2020}%
  \BibitemOpen
  \bibfield  {author} {\bibinfo {author} {\bibfnamefont {Y.}~\bibnamefont
  {Chen}}, \bibinfo {author} {\bibfnamefont {J.}~\bibnamefont {Gaudet}},
  \bibinfo {author} {\bibfnamefont {S.}~\bibnamefont {Dasgupta}}, \bibinfo
  {author} {\bibfnamefont {G.~G.}\ \bibnamefont {Marcus}}, \bibinfo {author}
  {\bibfnamefont {J.}~\bibnamefont {Lin}}, \bibinfo {author} {\bibfnamefont
  {T.}~\bibnamefont {Chen}}, \bibinfo {author} {\bibfnamefont {T.}~\bibnamefont
  {Tomita}}, \bibinfo {author} {\bibfnamefont {M.}~\bibnamefont {Ikhlas}},
  \bibinfo {author} {\bibfnamefont {Y.}~\bibnamefont {Zhao}}, \bibinfo {author}
  {\bibfnamefont {W.~C.}\ \bibnamefont {Chen}}, \bibinfo {author}
  {\bibfnamefont {M.~B.}\ \bibnamefont {Stone}}, \bibinfo {author}
  {\bibfnamefont {O.}~\bibnamefont {Tchernyshyov}}, \bibinfo {author}
  {\bibfnamefont {S.}~\bibnamefont {Nakatsuji}}, \ and\ \bibinfo {author}
  {\bibfnamefont {C.}~\bibnamefont {Broholm}},\ }\href {\doibase
  10.1103/PhysRevB.102.054403} {\bibfield  {journal} {\bibinfo  {journal}
  {Phys. Rev. B}\ }\textbf {\bibinfo {volume} {102}},\ \bibinfo {pages}
  {054403} (\bibinfo {year} {2020})}\BibitemShut {NoStop}%
\bibitem [{\citenamefont {Soh}\ \emph {et~al.}(2020)\citenamefont {Soh},
  \citenamefont {de~Juan}, \citenamefont {Qureshi}, \citenamefont {Jacobsen},
  \citenamefont {Wang}, \citenamefont {Guo},\ and\ \citenamefont
  {Boothroyd}}]{soh2020}%
  \BibitemOpen
  \bibfield  {author} {\bibinfo {author} {\bibfnamefont {J.-R.}\ \bibnamefont
  {Soh}}, \bibinfo {author} {\bibfnamefont {F.}~\bibnamefont {de~Juan}},
  \bibinfo {author} {\bibfnamefont {N.}~\bibnamefont {Qureshi}}, \bibinfo
  {author} {\bibfnamefont {H.}~\bibnamefont {Jacobsen}}, \bibinfo {author}
  {\bibfnamefont {H.-Y.}\ \bibnamefont {Wang}}, \bibinfo {author}
  {\bibfnamefont {Y.-F.}\ \bibnamefont {Guo}}, \ and\ \bibinfo {author}
  {\bibfnamefont {A.~T.}\ \bibnamefont {Boothroyd}},\ }\href {\doibase
  10.1103/PhysRevB.101.140411} {\bibfield  {journal} {\bibinfo  {journal}
  {Phys. Rev. B}\ }\textbf {\bibinfo {volume} {101}},\ \bibinfo {pages}
  {140411} (\bibinfo {year} {2020})}\BibitemShut {NoStop}%
\bibitem [{\citenamefont {Yamane}\ \emph {et~al.}(2017)\citenamefont {Yamane},
  \citenamefont {Gomonay}, \citenamefont {Velkov},\ and\ \citenamefont
  {Sinova}}]{yamane17}%
  \BibitemOpen
  \bibfield  {author} {\bibinfo {author} {\bibfnamefont {Y.}~\bibnamefont
  {Yamane}}, \bibinfo {author} {\bibfnamefont {O.}~\bibnamefont {Gomonay}},
  \bibinfo {author} {\bibfnamefont {H.}~\bibnamefont {Velkov}}, \ and\ \bibinfo
  {author} {\bibfnamefont {J.}~\bibnamefont {Sinova}},\ }\href {\doibase
  10.1103/PhysRevB.96.064408} {\bibfield  {journal} {\bibinfo  {journal} {Phys.
  Rev. B}\ }\textbf {\bibinfo {volume} {96}},\ \bibinfo {pages} {064408}
  (\bibinfo {year} {2017})}\BibitemShut {NoStop}%
\bibitem [{\citenamefont {Tveten}\ \emph {et~al.}(2013)\citenamefont {Tveten},
  \citenamefont {Qaiumzadeh}, \citenamefont {Tretiakov},\ and\ \citenamefont
  {Brataas}}]{tveten13}%
  \BibitemOpen
  \bibfield  {author} {\bibinfo {author} {\bibfnamefont {E.~G.}\ \bibnamefont
  {Tveten}}, \bibinfo {author} {\bibfnamefont {A.}~\bibnamefont {Qaiumzadeh}},
  \bibinfo {author} {\bibfnamefont {O.~A.}\ \bibnamefont {Tretiakov}}, \ and\
  \bibinfo {author} {\bibfnamefont {A.}~\bibnamefont {Brataas}},\ }\href
  {\doibase 10.1103/PhysRevLett.110.127208} {\bibfield  {journal} {\bibinfo
  {journal} {Phys. Rev. Lett.}\ }\textbf {\bibinfo {volume} {110}},\ \bibinfo
  {pages} {127208} (\bibinfo {year} {2013})}\BibitemShut {NoStop}%
\bibitem [{\citenamefont {Tserkovnyak}\ and\ \citenamefont
  {Xiao}(2018)}]{YaroslavPRL2018}%
  \BibitemOpen
  \bibfield  {author} {\bibinfo {author} {\bibfnamefont {Y.}~\bibnamefont
  {Tserkovnyak}}\ and\ \bibinfo {author} {\bibfnamefont {J.}~\bibnamefont
  {Xiao}},\ }\href {\doibase 10.1103/PhysRevLett.121.127701} {\bibfield
  {journal} {\bibinfo  {journal} {Phys. Rev. Lett.}\ }\textbf {\bibinfo
  {volume} {121}},\ \bibinfo {pages} {127701} (\bibinfo {year}
  {2018})}\BibitemShut {NoStop}%
\bibitem [{\citenamefont {Jones}\ \emph {et~al.}(2020)\citenamefont {Jones},
  \citenamefont {Zou}, \citenamefont {Zhang},\ and\ \citenamefont
  {Tserkovnyak}}]{jones2020}%
  \BibitemOpen
  \bibfield  {author} {\bibinfo {author} {\bibfnamefont {D.}~\bibnamefont
  {Jones}}, \bibinfo {author} {\bibfnamefont {J.}~\bibnamefont {Zou}}, \bibinfo
  {author} {\bibfnamefont {S.}~\bibnamefont {Zhang}}, \ and\ \bibinfo {author}
  {\bibfnamefont {Y.}~\bibnamefont {Tserkovnyak}},\ }\href {\doibase
  10.1103/PhysRevB.102.140411} {\bibfield  {journal} {\bibinfo  {journal}
  {Phys. Rev. B}\ }\textbf {\bibinfo {volume} {102}},\ \bibinfo {pages}
  {140411} (\bibinfo {year} {2020})}\BibitemShut {NoStop}%
\bibitem [{\citenamefont {Kim}\ and\ \citenamefont {Chung}(2021)}]{KimSci2021}%
  \BibitemOpen
  \bibfield  {author} {\bibinfo {author} {\bibfnamefont {S.~K.}\ \bibnamefont
  {Kim}}\ and\ \bibinfo {author} {\bibfnamefont {S.~B.}\ \bibnamefont
  {Chung}},\ }\href {\doibase 10.21468/SciPostPhys.10.3.068} {\bibfield
  {journal} {\bibinfo  {journal} {SciPost Phys.}\ }\textbf {\bibinfo {volume}
  {10}},\ \bibinfo {pages} {68} (\bibinfo {year} {2021})}\BibitemShut {NoStop}%
\bibitem [{\citenamefont {Zou}\ \emph {et~al.}(2019)\citenamefont {Zou},
  \citenamefont {Kim},\ and\ \citenamefont {Tserkovnyak}}]{Zou2019}%
  \BibitemOpen
  \bibfield  {author} {\bibinfo {author} {\bibfnamefont {J.}~\bibnamefont
  {Zou}}, \bibinfo {author} {\bibfnamefont {S.~K.}\ \bibnamefont {Kim}}, \ and\
  \bibinfo {author} {\bibfnamefont {Y.}~\bibnamefont {Tserkovnyak}},\ }\href
  {\doibase 10.1103/PhysRevB.99.180402} {\bibfield  {journal} {\bibinfo
  {journal} {Phys. Rev. B}\ }\textbf {\bibinfo {volume} {99}},\ \bibinfo
  {pages} {180402} (\bibinfo {year} {2019})}\BibitemShut {NoStop}%
\bibitem [{\citenamefont {Tserkovnyak}\ \emph {et~al.}(2020)\citenamefont
  {Tserkovnyak}, \citenamefont {Zou}, \citenamefont {Kim},\ and\ \citenamefont
  {Takei}}]{YaroslavPRB2020}%
  \BibitemOpen
  \bibfield  {author} {\bibinfo {author} {\bibfnamefont {Y.}~\bibnamefont
  {Tserkovnyak}}, \bibinfo {author} {\bibfnamefont {J.}~\bibnamefont {Zou}},
  \bibinfo {author} {\bibfnamefont {S.~K.}\ \bibnamefont {Kim}}, \ and\
  \bibinfo {author} {\bibfnamefont {S.}~\bibnamefont {Takei}},\ }\href
  {\doibase 10.1103/PhysRevB.102.224433} {\bibfield  {journal} {\bibinfo
  {journal} {Phys. Rev. B}\ }\textbf {\bibinfo {volume} {102}},\ \bibinfo
  {pages} {224433} (\bibinfo {year} {2020})}\BibitemShut {NoStop}%
\bibitem [{\citenamefont {Cheng}\ \emph {et~al.}(2014)\citenamefont {Cheng},
  \citenamefont {Xiao}, \citenamefont {Niu},\ and\ \citenamefont
  {Brataas}}]{cheng2014}%
  \BibitemOpen
  \bibfield  {author} {\bibinfo {author} {\bibfnamefont {R.}~\bibnamefont
  {Cheng}}, \bibinfo {author} {\bibfnamefont {J.}~\bibnamefont {Xiao}},
  \bibinfo {author} {\bibfnamefont {Q.}~\bibnamefont {Niu}}, \ and\ \bibinfo
  {author} {\bibfnamefont {A.}~\bibnamefont {Brataas}},\ }\href {\doibase
  10.1103/PhysRevLett.113.057601} {\bibfield  {journal} {\bibinfo  {journal}
  {Phys. Rev. Lett.}\ }\textbf {\bibinfo {volume} {113}},\ \bibinfo {pages}
  {057601} (\bibinfo {year} {2014})}\BibitemShut {NoStop}%
\bibitem [{\citenamefont {Zhang}\ and\ \citenamefont
  {Zhang}(2009)}]{shulei2009}%
  \BibitemOpen
  \bibfield  {author} {\bibinfo {author} {\bibfnamefont {S.}~\bibnamefont
  {Zhang}}\ and\ \bibinfo {author} {\bibfnamefont {S.~S.-L.}\ \bibnamefont
  {Zhang}},\ }\href {\doibase 10.1103/PhysRevLett.102.086601} {\bibfield
  {journal} {\bibinfo  {journal} {Phys. Rev. Lett.}\ }\textbf {\bibinfo
  {volume} {102}},\ \bibinfo {pages} {086601} (\bibinfo {year}
  {2009})}\BibitemShut {NoStop}%
\bibitem [{\citenamefont {Tatara}\ \emph {et~al.}(2008)\citenamefont {Tatara},
  \citenamefont {Kohno},\ and\ \citenamefont {Shibata}}]{tatara08}%
  \BibitemOpen
  \bibfield  {author} {\bibinfo {author} {\bibfnamefont {G.}~\bibnamefont
  {Tatara}}, \bibinfo {author} {\bibfnamefont {H.}~\bibnamefont {Kohno}}, \
  and\ \bibinfo {author} {\bibfnamefont {J.}~\bibnamefont {Shibata}},\ }\href
  {\doibase https://doi.org/10.1016/j.physrep.2008.07.003} {\bibfield
  {journal} {\bibinfo  {journal} {Physics Reports}\ }\textbf {\bibinfo {volume}
  {468}},\ \bibinfo {pages} {213} (\bibinfo {year} {2008})}\BibitemShut
  {NoStop}%
\bibitem [{\citenamefont {Swaving}\ and\ \citenamefont
  {Duine}(2011)}]{swaving11}%
  \BibitemOpen
  \bibfield  {author} {\bibinfo {author} {\bibfnamefont {A.~C.}\ \bibnamefont
  {Swaving}}\ and\ \bibinfo {author} {\bibfnamefont {R.~A.}\ \bibnamefont
  {Duine}},\ }\href {\doibase 10.1103/PhysRevB.83.054428} {\bibfield  {journal}
  {\bibinfo  {journal} {Phys. Rev. B}\ }\textbf {\bibinfo {volume} {83}},\
  \bibinfo {pages} {054428} (\bibinfo {year} {2011})}\BibitemShut {NoStop}%
\bibitem [{\citenamefont {Yuan}\ \emph {et~al.}(2018)\citenamefont {Yuan},
  \citenamefont {Wang}, \citenamefont {Yung},\ and\ \citenamefont
  {Wang}}]{Yuan2018}%
  \BibitemOpen
  \bibfield  {author} {\bibinfo {author} {\bibfnamefont {H.~Y.}\ \bibnamefont
  {Yuan}}, \bibinfo {author} {\bibfnamefont {W.}~\bibnamefont {Wang}}, \bibinfo
  {author} {\bibfnamefont {M.-H.}\ \bibnamefont {Yung}}, \ and\ \bibinfo
  {author} {\bibfnamefont {X.~R.}\ \bibnamefont {Wang}},\ }\href {\doibase
  10.1103/PhysRevB.97.214434} {\bibfield  {journal} {\bibinfo  {journal} {Phys.
  Rev. B}\ }\textbf {\bibinfo {volume} {97}},\ \bibinfo {pages} {214434}
  (\bibinfo {year} {2018})}\BibitemShut {NoStop}%
\bibitem [{\citenamefont {Bar{\textquotesingle}yakhtar}\ \emph
  {et~al.}(1985)\citenamefont {Bar{\textquotesingle}yakhtar}, \citenamefont
  {Ivanov},\ and\ \citenamefont {Chetkin}}]{Bar_yakhtar_1985}%
  \BibitemOpen
  \bibfield  {author} {\bibinfo {author} {\bibfnamefont {V.~G.}\ \bibnamefont
  {Bar{\textquotesingle}yakhtar}}, \bibinfo {author} {\bibfnamefont {B.~A.}\
  \bibnamefont {Ivanov}}, \ and\ \bibinfo {author} {\bibfnamefont {M.~V.}\
  \bibnamefont {Chetkin}},\ }\href {\doibase 10.1070/pu1985v028n07abeh003871}
  {\bibfield  {journal} {\bibinfo  {journal} {Soviet Physics Uspekhi}\ }\textbf
  {\bibinfo {volume} {28}},\ \bibinfo {pages} {563} (\bibinfo {year}
  {1985})}\BibitemShut {NoStop}%
\bibitem [{\citenamefont {Galkina}\ and\ \citenamefont
  {Ivanov}(2018)}]{Galkina2018}%
  \BibitemOpen
  \bibfield  {author} {\bibinfo {author} {\bibfnamefont {E.~G.}\ \bibnamefont
  {Galkina}}\ and\ \bibinfo {author} {\bibfnamefont {B.~A.}\ \bibnamefont
  {Ivanov}},\ }\href {\doibase 10.1063/1.5041427} {\bibfield  {journal}
  {\bibinfo  {journal} {Low Temperature Physics}\ }\textbf {\bibinfo {volume}
  {44}},\ \bibinfo {pages} {618} (\bibinfo {year} {2018})}\BibitemShut
  {NoStop}%
\bibitem [{\citenamefont {Ivanov}\ and\ \citenamefont
  {Sheka}(1994)}]{ivanov94}%
  \BibitemOpen
  \bibfield  {author} {\bibinfo {author} {\bibfnamefont {B.~A.}\ \bibnamefont
  {Ivanov}}\ and\ \bibinfo {author} {\bibfnamefont {D.~D.}\ \bibnamefont
  {Sheka}},\ }\href {\doibase 10.1103/PhysRevLett.72.404} {\bibfield  {journal}
  {\bibinfo  {journal} {Phys. Rev. Lett.}\ }\textbf {\bibinfo {volume} {72}},\
  \bibinfo {pages} {404} (\bibinfo {year} {1994})}\BibitemShut {NoStop}%
\bibitem [{\citenamefont {Clarke}\ \emph {et~al.}(2008)\citenamefont {Clarke},
  \citenamefont {Tretiakov}, \citenamefont {Chern}, \citenamefont {Bazaliy},\
  and\ \citenamefont {Tchernyshyov}}]{clarke08}%
  \BibitemOpen
  \bibfield  {author} {\bibinfo {author} {\bibfnamefont {D.~J.}\ \bibnamefont
  {Clarke}}, \bibinfo {author} {\bibfnamefont {O.~A.}\ \bibnamefont
  {Tretiakov}}, \bibinfo {author} {\bibfnamefont {G.-W.}\ \bibnamefont
  {Chern}}, \bibinfo {author} {\bibfnamefont {Y.~B.}\ \bibnamefont {Bazaliy}},
  \ and\ \bibinfo {author} {\bibfnamefont {O.}~\bibnamefont {Tchernyshyov}},\
  }\href {\doibase 10.1103/PhysRevB.78.134412} {\bibfield  {journal} {\bibinfo
  {journal} {Phys. Rev. B}\ }\textbf {\bibinfo {volume} {78}},\ \bibinfo
  {pages} {134412} (\bibinfo {year} {2008})}\BibitemShut {NoStop}%
\bibitem [{\citenamefont {Pylypovskyi}\ \emph {et~al.}(2012)\citenamefont
  {Pylypovskyi}, \citenamefont {Sheka},\ and\ \citenamefont
  {Gaididei}}]{pylypovskyi2012}%
  \BibitemOpen
  \bibfield  {author} {\bibinfo {author} {\bibfnamefont {O.~V.}\ \bibnamefont
  {Pylypovskyi}}, \bibinfo {author} {\bibfnamefont {D.~D.}\ \bibnamefont
  {Sheka}}, \ and\ \bibinfo {author} {\bibfnamefont {Y.}~\bibnamefont
  {Gaididei}},\ }\href {\doibase 10.1103/PhysRevB.85.224401} {\bibfield
  {journal} {\bibinfo  {journal} {Phys. Rev. B}\ }\textbf {\bibinfo {volume}
  {85}},\ \bibinfo {pages} {224401} (\bibinfo {year} {2012})}\BibitemShut
  {NoStop}%
\end{thebibliography}%

\end{document}